\PassOptionsToPackage{unicode}{hyperref}
\PassOptionsToPackage{hyphens}{url}
\documentclass[conference]{IEEEtran}

\usepackage{dnxieeestyle}

\usepackage{xpatch}

\makeatletter
\patchcmd\blx@bblinput{\blx@blxinit}
                      {\blx@blxinit
                      }{}{\fail}
\makeatother

\title{Algebraic Vertex Ordering of a Sparse Graph
for \\ Adjacency Access Locality and Graph Compression}

\author{\IEEEauthorblockN{Dimitris~Floros\IEEEauthorrefmark{1}
    \qquad 
    Nikos~Pitsianis\IEEEauthorrefmark{2}\IEEEauthorrefmark{3}
      \qquad
    Xiaobai~Sun\IEEEauthorrefmark{3}}
    \\
    \IEEEauthorblockA{\footnotesize \begin{tabular}{c @{\qquad\qquad} c @{\qquad\qquad} c}
        \IEEEauthorrefmark{1}Nicholas~School~of~the~Environment
        &
        \IEEEauthorrefmark{2}Department~of~Electrical~and~Computer~Engineering
        &
        \IEEEauthorrefmark{3}Department~of~Computer~Science
        \\
        Duke~University
        &
        Aristotle~University~of~Thessaloniki
        &
        Duke~University
        \\
        Durham,~NC~27708,~USA
        &
        Thessaloniki~54124,~Greece
        &
        Durham,~NC~27708,~USA
      \end{tabular}}}

\begin{document}

\bstctlcite{IEEEexample:BSTcontrol}

\maketitle

\savebox{\tempbox}{
\begin{minipage}{\textwidth}    \setcaptype{figure}\centering
\begin{subfigure}{0.18\linewidth}
    \centering
    \includegraphics[width=\linewidth]{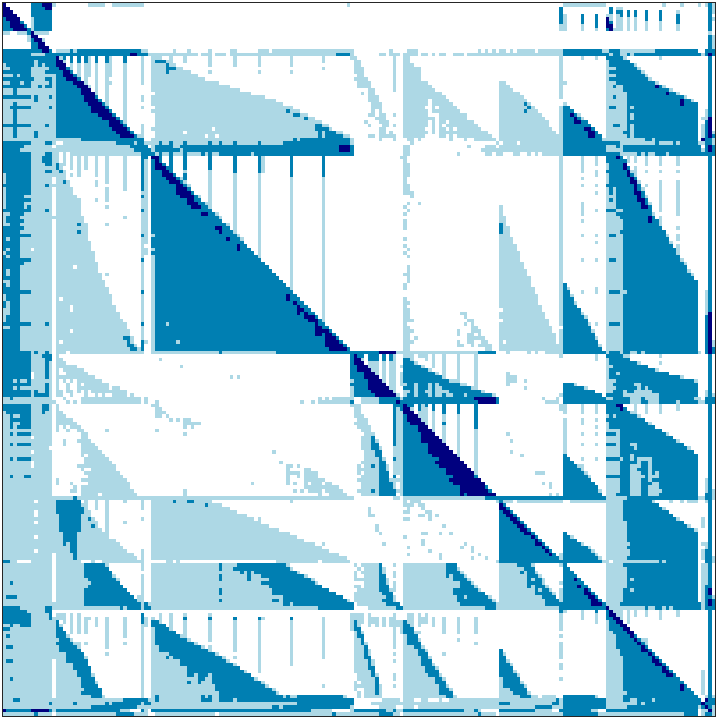}
    \caption{\small {\sc doi}}
  \end{subfigure}
\hfill
  \begin{subfigure}{0.18\linewidth}
    \centering
    \includegraphics[width=\linewidth]{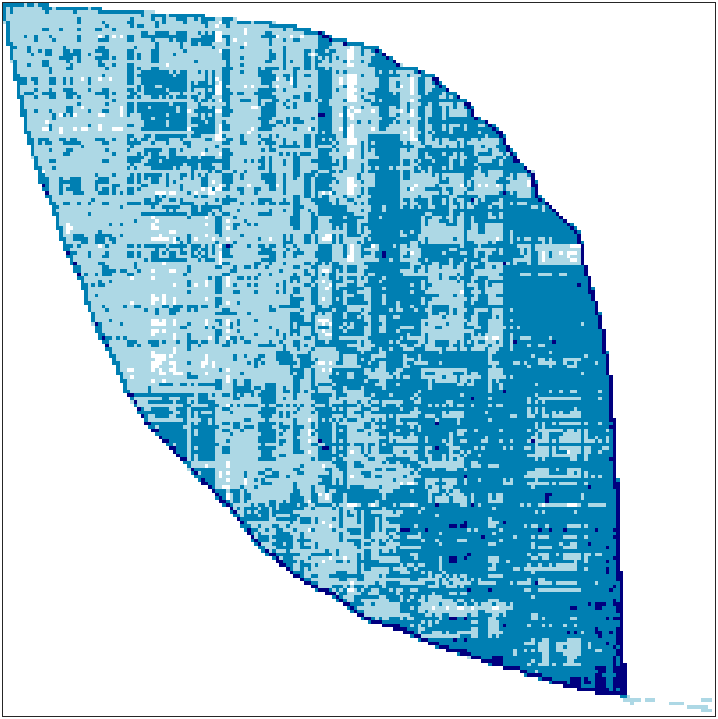}
    \caption{\small \rcm}
  \end{subfigure}
\hfill
  \begin{subfigure}{0.18\linewidth}
    \centering
    \includegraphics[width=\linewidth]{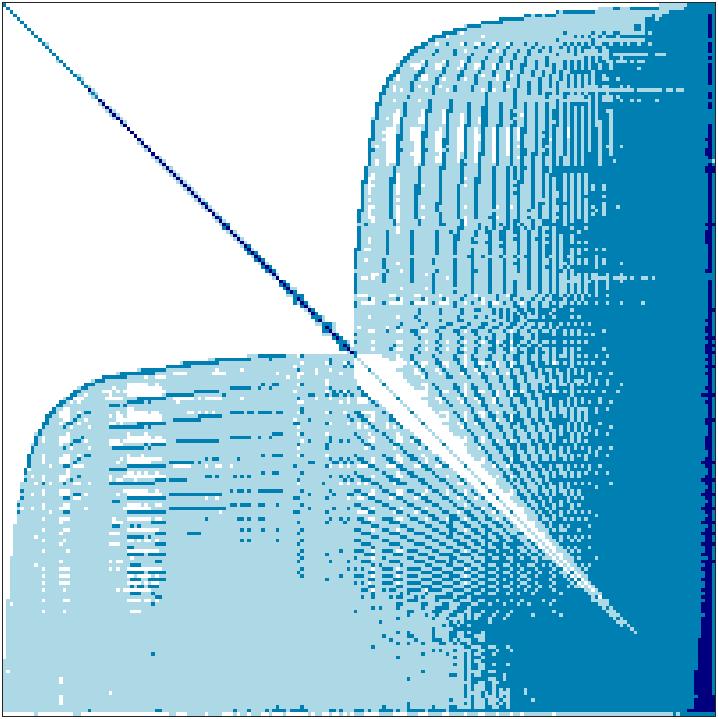}
    \caption{\small \slashburn}
  \end{subfigure}
\hfill
  \begin{subfigure}{0.18\linewidth}
    \centering
    \includegraphics[width=\linewidth]{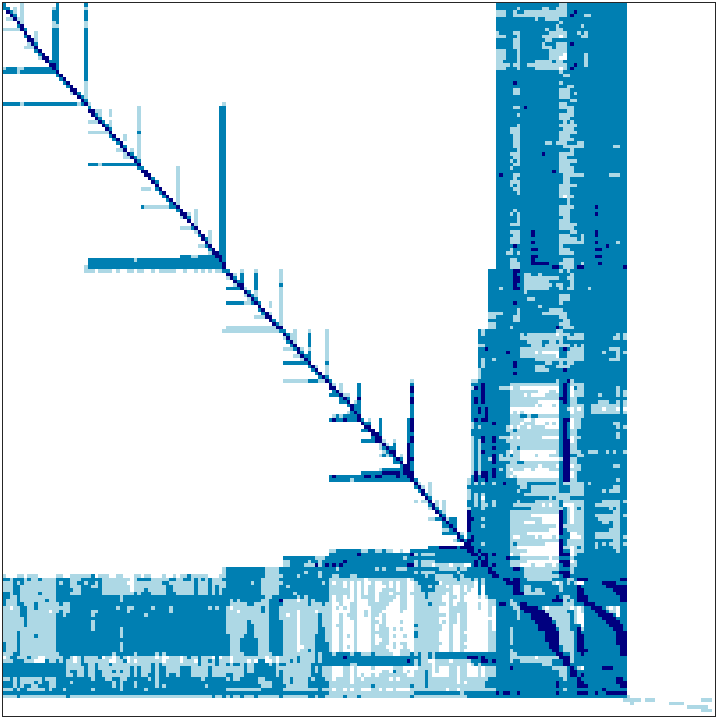}
    \caption{\small \amd}
  \end{subfigure}
\hfill
  \begin{subfigure}{0.18\linewidth}
    \centering
    \includegraphics[width=\linewidth]{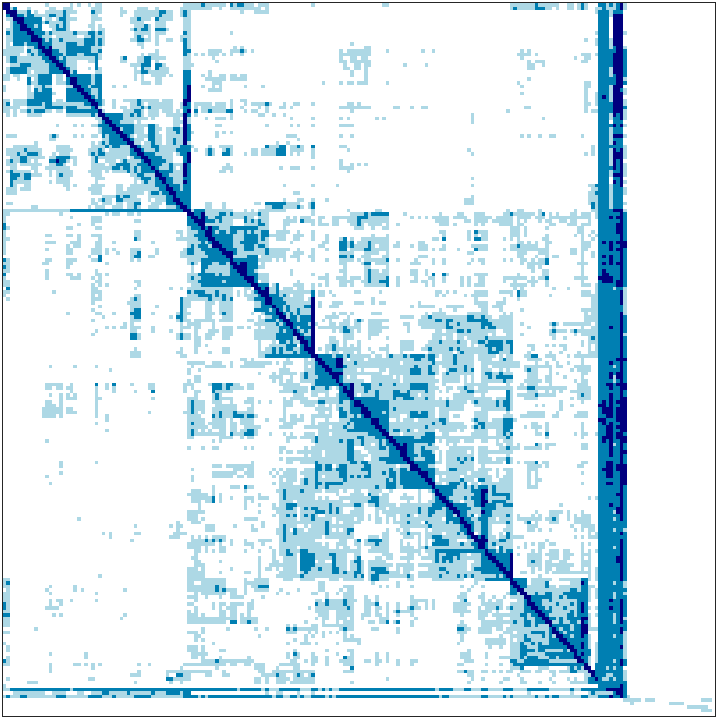}
    \caption{\small \vifps}
  \end{subfigure}
\caption{ The adjacency-matrix images of the citation graph
    $\aps \!=\! G(V,E)$ in five different vertex
    orderings. The graph is a representation of the American Physical
    Society ({\sf APS}) publication up to the year $2020$, with
    $|V|\!=\! \num{667365}$ articles and $|E| \!=\! \num{8849630}$
    citation links~\cite{aps-2020}. Each image pixel $(i,j)$ represents a citation
    subgraph ${\footnotesize G(V_i,V_j,E_{ij})}$ with
    ${\footnotesize |V_i|\!=\!|V_j|\!=\!\num{3336}}$,
    $E_{ij}=(V_i\times V_j)\cap E$.  A darker pixel indicates a denser
    subgraph. The five ordering methods are: (a) the lexicographical
    order of the digital object identifiers (\text{DOIs}) for the {\sf
      APS} articles, (b) the reverse Cuthill-McKee (\rcm{})
    method~\cite{cuthill1969reducingbandwidth}, (c) the \slashburn{}
    method~\cite{lim2014slashburngraph}, (d) the approximate minimum
    degree (\amd{}) method~\cite{amestoy1996approximateminimum}, and
    (e) the new method \vifps{}. The properties and performance
    assessment of these orderings with regard to graph compression are
    elaborated in the rest of the paper.  }  \label{fig:teaser-aps2020}
  \vspace*{-1em}
\end{minipage}

 } 

\begin{figure}[t]
  \rlap{\usebox\tempbox}
\end{figure}
\afterpage{
  \begin{figure}[t]\rule{0pt}{\dimexpr \ht\tempbox+\dp\tempbox}
  \end{figure}
}

\begin{abstract}
  In this work, we establish theoretical and practical connections
between vertex indexing for sparse graph/network compression and
matrix ordering for sparse matrix-vector multiplication and variable
elimination. We present a fundamental analysis of adjacency access
locality in vertex ordering from the perspective of graph composition
of, or decomposition into, elementary compact graphs. We introduce an
algebraic indexing approach that maintains the advantageous features
of existing methods, mitigates their shortcomings, and adapts to the
degree distribution. The new method demonstrates superior and
versatile performance in graph compression across diverse types of
graphs. It also renders proportional improvement in the efficiency of
matrix-vector multiplications for subspace iterations in response to
random walk queries on a large network.

 \end{abstract}


\section{Introduction}
\label{sec:introduction}

In a modern data, knowledge, or information system, the datum entities
(or feature vectors) are typically linked by a direct or induced
pairwise adjacency relationship and represented as the vertices of a
big graph/network $G(V,E)$ with vertex set $V$ and edge set $E$.  The
edge set $E\subset V\times V$ represents the pairwise adjacency
relation.  The edges may be undirected or directed.
The large graph is usually sparse and structured compared to
Erd\H{o}s-R\'{e}nyi random graphs of the same sparsity.  For
data-information management and processing, the vertices/nodes are
indexed from $1$ to $n=|V|$ by a $1$-to-$1$ mapping from the datum
labels or identification numbers.
How the vertices are sequentially indexed or ordered significantly
impacts the space-time performance of the system in two related key
aspects---the graph compression beyond the simple exclusion of absent
links and the efficiency in response to frequent adjacency queries for
retrieval and referral while the graph is accessed via a compressed
representation. In this paper, we present our study and findings on
vertex ordering for graph compression.

A graph compression maintaining real-time adjacency accesses (ideally
in memory) is essentially an integral compression of individual
adjacency lists on the graph with small compression/decompression
windows.
Every vertex $v$ has an adjacency/neighbor list $\mathcal{N}(v)$
consisting of its incident edges or, in interchangeable terms, its
adjacent nodes or immediate neighbors. With a vertex ordering, the
adjacency list is represented by a subsequence of $\{1,2,\dots,n\}$.
The degree of node $v$, $d(v)$, is the number of its neighbors, $d(v)
= |\mathcal{N}(v)|$. If graph $G$ is directed, $\mathcal{N}(v)$ has
two sublists: $\mathcal{N}_{\rm in}(v)$ of the incoming links and
$\mathcal{N}_{\rm out}(v)$ of the outgoing links. Invariant to vertex
ordering, the total number of adjacency lists is $n$; the total number
of the list items is $m=|E|$. Nonetheless, the integer subsequences
representing the adjacency lists change from one vertex ordering to
another, their compression is thereby affected first and foremost by
vertex ordering.

We also use $\mathcal{N}(v)$ to denote the $1$-hop neighborhood graph
centered at vertex $v$, it is a topological feature of vertex $v$ on
graph $G$.  When topologically more similar or overlapped nodes are
indexed closer to each other, the vertex ordering provides greater
room for subsequent compression of the adjacency lists.  Such ordering
also potentially improves the spatial localities in overall adjacency
accesses and, thereby, the efficiency in query response.
Alternatively, a vertex ordering that increases the spatial localities
in adjacency lists benefits graph compression. Briefly, grouping
similar adjacency lists or minimizing their index distances is the key
ingredient in an effective vertex ordering scheme.

There are several notable vertex ordering schemes for the compression
of data-information graphs subject to the condition of maintaining
real-time adjacency
accesses~\cite{dhulipala2016compressinggraphs,besta2019surveytaxonomy}.
Some of them are specific to certain types of networks/graphs,
followed by customized list compression techniques, or both.
For instance, for web-graph compression, Silvestri suggested ordering
the webpages by the lexicographic ordering of their uniform resource
locator~({\sc url}) addresses~\cite{silvestri2007sortingout}. The
heuristic is that the similarity in adjacency patterns is well
correlated with the {\sc url} proximity.
Boldi and Vigna~(BV) introduced a novel compression
technique~\cite{boldi2004webgraphframework}.  Their main idea is to
encode the variation/gap in the neighborhood pattern of an adjacency
list $\mathcal{N}(i)$ from a local prototype pattern among the
adjacency lists $\mathcal{N}(j)$ within a small index block, say, $8$
consecutive indices per block.  The BV approach achieved a remarkable
web-graph compression, \SI{3}{\textrm{bits}} per web link, which was
further reduced to \SI{2}{\textrm{bits}} per link by
others~\cite{besta2019surveytaxonomy}.

For social-network compression, Chierichetti, Kumar, Lattanzi et
al.~({\sc ckl}) advocated the use of the \shingle{} scores and
ordering scheme, which were introduced by Gibson, Kumar, and
Tomkins~\cite{chierichetti2009compressingsocial,gibson2005discovering}.
The \shingle{} scores measure and encode the overlap/similarity
among the $1$-hop neighborhood graphs $\mathcal{N}(v)$, $v\in V$, and
are subsequently used to order the vertices with a hashing function.
We can expect the \shingle{} scores to be extended
straightforwardly to $h$-hop neighborhood graphs $\mathcal{N}_{h}(v)$
with $h>1$.  Social networks tend to follow the power law in their
degree distributions due to the preferential attachment as modeled or
stylized by Barabasi and
Albert~(BA)~\cite{barabasi1999}.  A small
number of nodes on a BA or BA-like network have very high degrees,
their overreaching presence in many neighborhood graphs reduces the
differentiation power by the \shingle{} scores.  To overcome this
problem, the {\sc ckl} approach removes the nodes with high degrees,
above $d_{\tau}$, before ordering the other nodes by their
neighborhood similarity in the remaining graph. The parameter
$d_{\tau}$ is set above the average degree, and it can be determined
from the degree distribution.

For BA-like networks/graphs, not necessarily social networks, an
ordering scheme named \slashburn{} was developed by Lim, Kang, and
Faloutsos~\cite{lim2014slashburngraph}.  It first removes the
top-$\ell$ high-degree nodes from the current graph. The removal
criterion may be understood as the alternative to the {\sc ckl}
criterion. \slashburn{} then decomposes the remaining graph into
its connected components~(CCs).  The node-removal and CC-decomposition
steps are applied to each of the connected components recursively.
We may view \slashburn{} extracting and expressing adjacency
localities as layered connected components, not limited to
neighborhood graphs with a fixed hop length as with the {\sc ckl}
approach.  \slashburn{} was shown superior to \shingle{}, by
certain graph compression measures, in a benchmarking study with
BA-like networks.

Interestingly, the Cuthill-McKee~({\sc cm}) ordering or its reversal
(\rcm{}), and the Fiedler-spectral ordering, well known for their
roles in sparse matrix
computation~\cite{cuthill1969reducingbandwidth,george1981computersolution},
are frequently used as base cases for performance comparisons among
vertex ordering schemes for graph compression.
Our first finding is that \slashburn{} is outperformed, by and large,
on BA-like graphs by \amd{} and \ned{}, which are among the most
effective ordering schemes for space-time efficient variable
elimination~\cite{amestoy2004algorithm837,amestoy1996approximateminimum,
marx2021proceedings2021,karypis1998fasthigha}, although not obviously
relevant to graph compression as \rcm{}.

\begin{figure}
  \vspace{-1em}   
  \centering
  \hspace{1em}
  \begin{subfigure}{0.27\linewidth}
    \includegraphics[width=\linewidth]{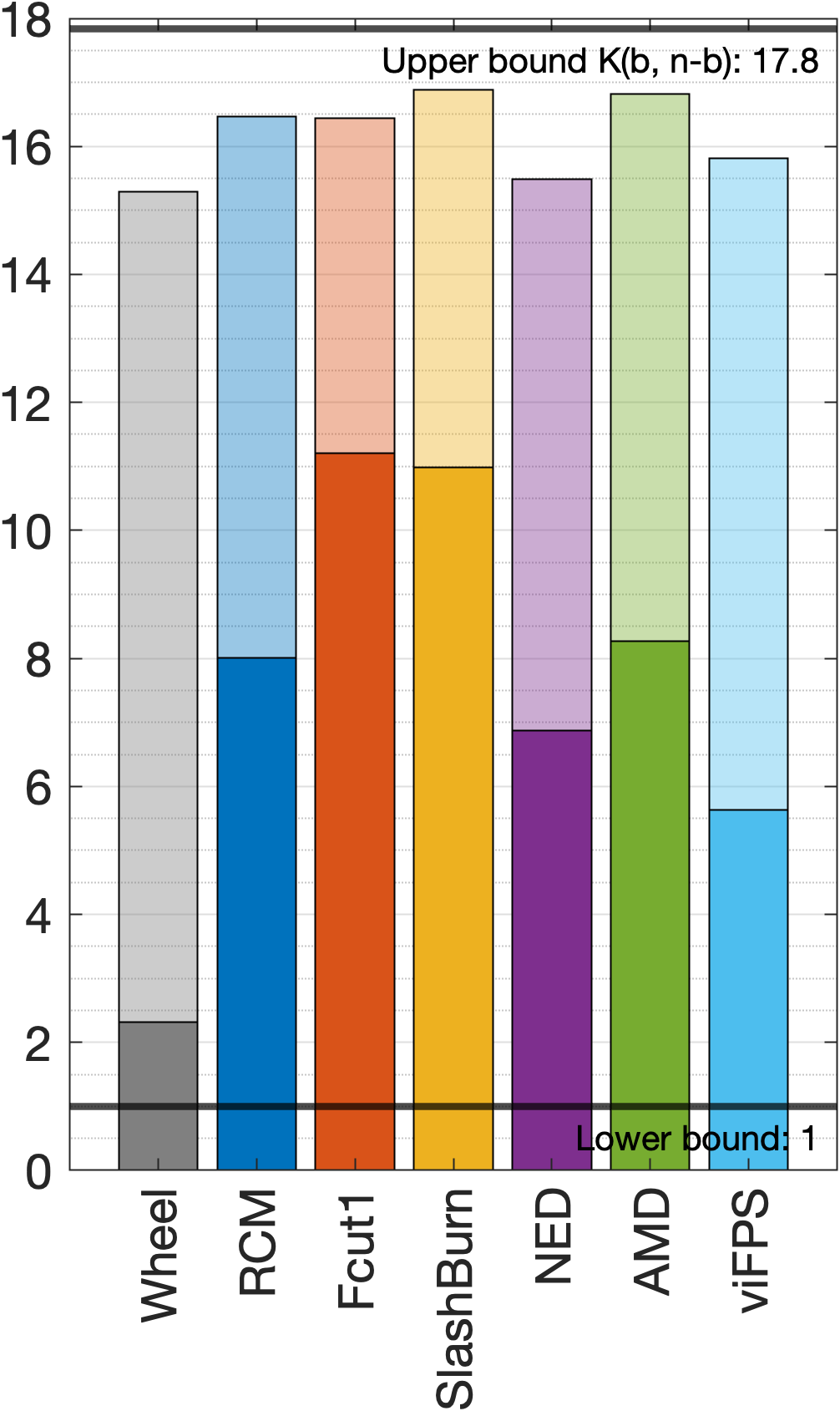}
    \subcaption{\aps{}}
  \end{subfigure}
  \hspace{1em}
  \begin{subfigure}{0.27\linewidth}
    \includegraphics[width=\linewidth]{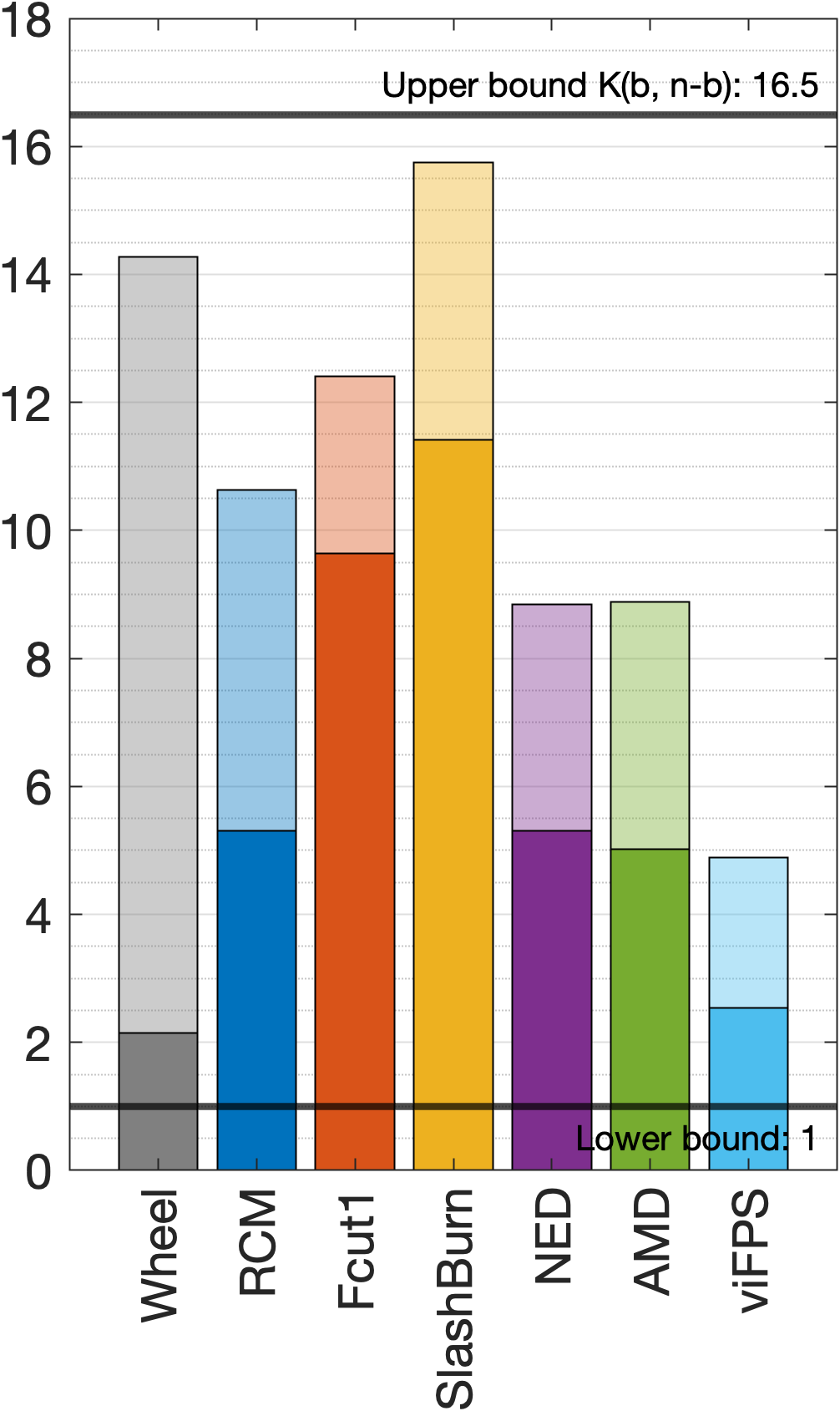}
    \subcaption{$G_{\rm ws}$}
  \end{subfigure}
  \hspace{1em}
  \begin{subfigure}{0.27\linewidth}
    \includegraphics[width=\linewidth]{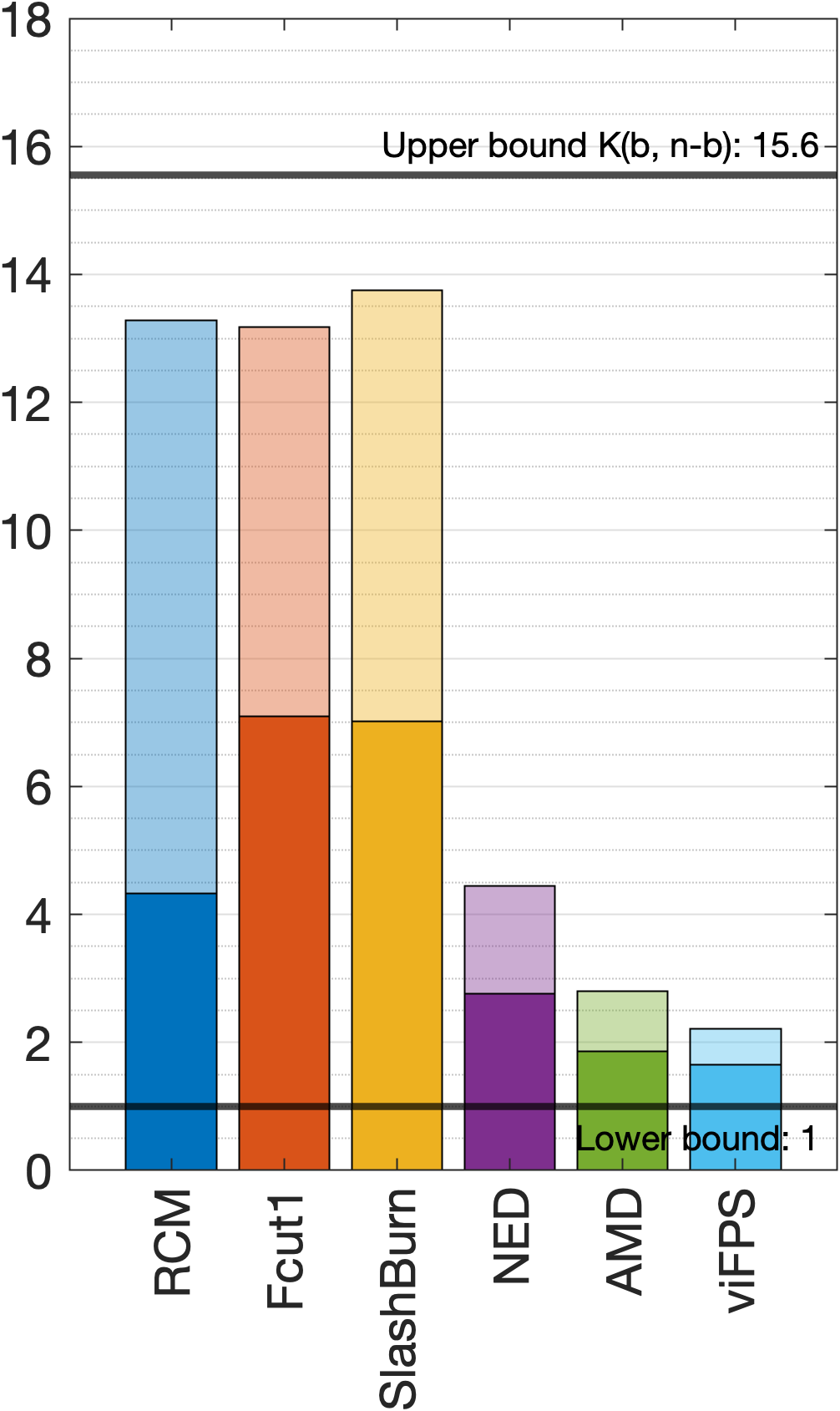}
    \subcaption{$T_{\rm binomial}$}
  \end{subfigure}
  \hspace{1em}
\vspace{-.5em}   
  \caption{Pictorial description of the entries of Table~\ref{tab:schemes-orders}
    for three graphs---\aps{}, $G_{\rm ws}$, and
    $T_{\rm binomial}$.  The bars in solid colors represent the
    $\mloggapa(G,\pi)$ scores: the lower, the better. They are
    between the lower and upper bounds in black lines by (\ref{eq:mLogGapA-lower-bound})
    and (\ref{eq:mLogGapA-upper-bound}).  The segments in lighter colors represent the
    $\Delta(G,\pi)$ values as structure indicators: a short
    light-colored segment over a short dark-colored bar indicates that
    more neighbors are placed within a short range on average.  The
    grey bars represent the reference values in (\ref{eq:warning-condition}) and
    (\ref{eq:wheel-mLogGapA-min}). It is a practical success when an ordering reaches
    close to the solid grey value on a non-elementary graph.
  } \label{fig:aal-measures}
  \vspace{-1em} 
\end{figure}

Our study of vertex ordering is comprehensive in two aspects:
\begin{inparaenum}[(a)]
\item we consider diverse types of sparse graphs, and
\item we analyze the match/mismatch between well-recognized graph
  types and popularly used ordering schemes in the context of
  adjacency access locality and graph compression.
\end{inparaenum}
We establish theoretical and practical connections between vertex
indexing for sparse graph/network compression and matrix ordering for
sparse matrix-vector multiplication and variable elimination.  These
connections led us to the first comparison between \slashburn{} and
\amd{}.

With this work, we make two main contributions.  First, we present a
fundamental analysis of adjacency access locality (\aal{}) in
vertex ordering from the perspective of graph/matrix composition of
elementary compact graphs/matrices.  The analysis is instrumental to
vertex ordering studies in more than one aspect.  It enables us to
identify the lower and upper bounds on the feasible \aal{} scores,
by the measures in (\ref{eq:mLogA}) and (\ref{eq:mLogGapA}), and,
thereby, explore previously unknown potentials or limitations.  We
provide additional reference \aal{} values not only for assessing
the performance of a vertex ordering but also for inferring the
substructures captured by the ordering. Such a frame of reference was
absent in previous performance assessments. More importantly, the
analysis sheds light on new ways for obtaining better approximate
solutions to the NP-hard vertex ordering problem.

Next, we present \vifps{}, a versatile indexing method for compression
of diverse types of graphs.  We maintain the advantageous features in
the state-of-the-art indexing schemes and mitigate their shortcomings.
The new method makes recursive {\em Fiedler} partition and ordering
conditioned by what we refer to as the {\em Pareto Splits} in
adaptation to the graph-degree statistics.
We show in Sections~\ref{sec:existing-limitations}
and~\ref{sec:viFPS-description} the outstanding performance of
\vifps{} on diverse types of graphs, competitive or superior to the
state-of-the-art ordering schemes for each type of graph.
As expected, the new method benefits sparse matrix computation as
well.  We show, in particular, improved efficiency for subspace
iterations in responding to random-walk queries on a large and
sparse network.

\section{Adjacency access locality analysis}
\label{sec:aal-analysis}

\subsection{Measures}
\label{sec:aal-measures}

\begin{figure}
  \centering
  \hspace*{1em}  
  \begin{subfigure}{0.30\linewidth}
    \includegraphics[width=\linewidth]{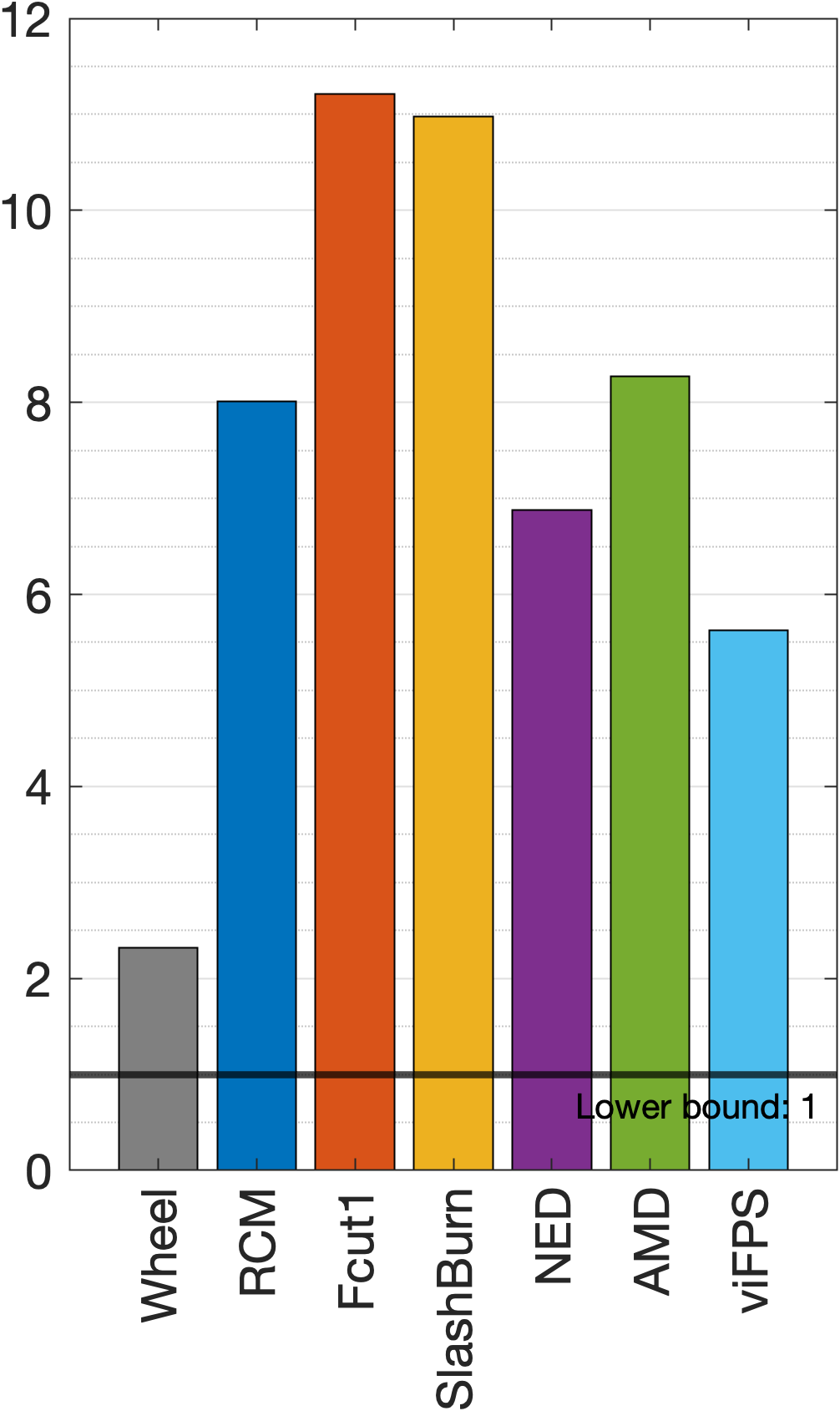}
    \subcaption{\mloggapa}
  \end{subfigure}
\hspace*{2.5em}
\begin{subfigure}{0.30\linewidth}
    \includegraphics[width=\linewidth]{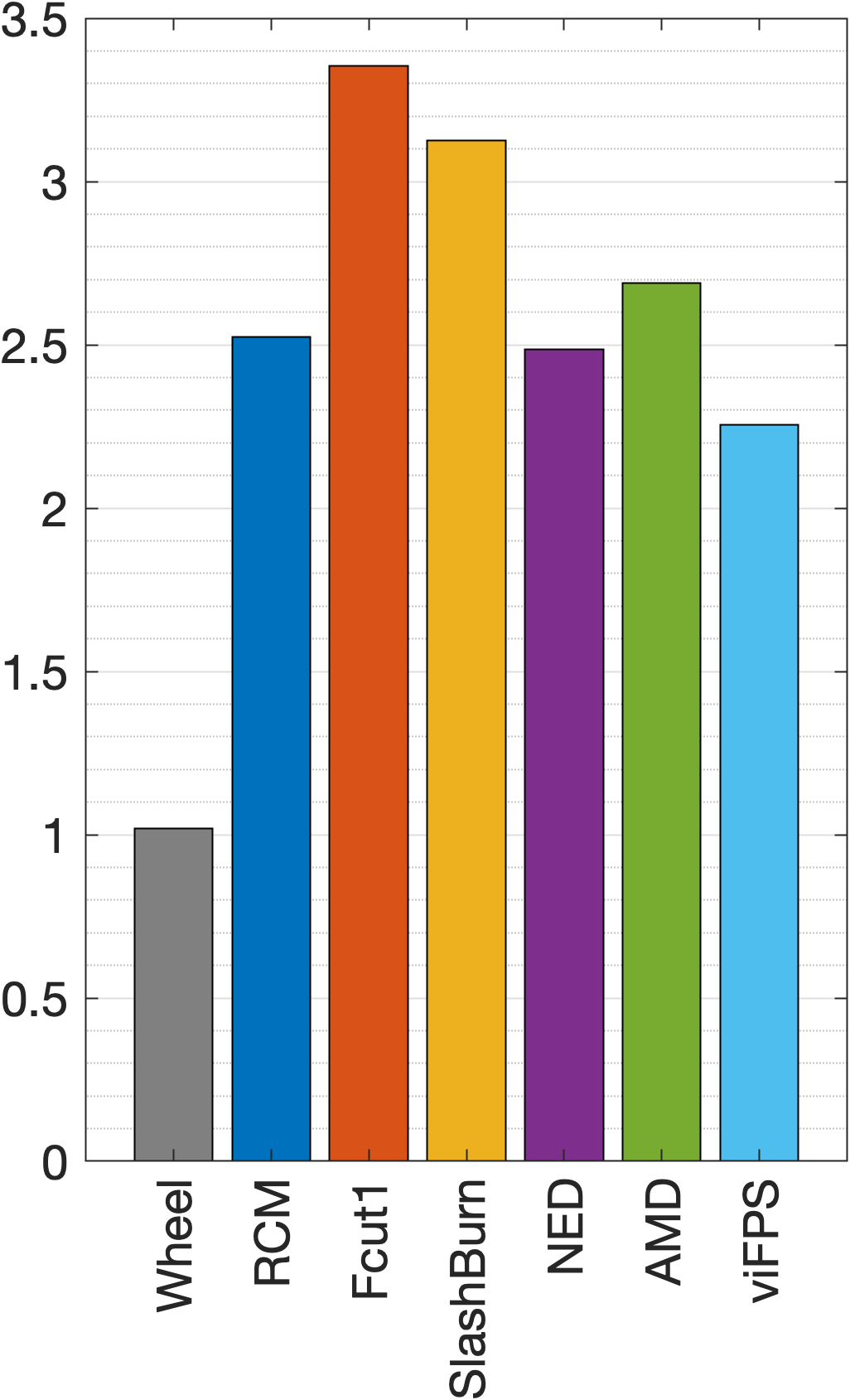}
    \subcaption{file sizes in HDF5}
  \end{subfigure}
  \hspace*{1em}
\vspace{-0.5em}
  \caption{A demonstration of the close correlation between
    \mloggapa{} of (\ref{eq:mLogGapA}) on vertex ordering and {\sf
      HDF5} for general-purpose compression. {\bf Left:} the
    \mloggapa{} scores of six different orderings on the citation
    graph \aps; {\bf Right:} the corresponding file
    sizes in {\sf HDF5} format. The file sizes are normalized by that
    for the reference graph $G_{\rm wheel}$ of the same size and
    sparsity in the \amd{} ordering, based on (\ref{eq:warning-condition}).  }
  \label{fig:mLogGapA-hdf5-correlation}
  \vspace{-2em} 
\end{figure}

We adopt two well-established measures of vertex orderings for graph
compression, with a slight modification.  Denote by $G(V,E)$ the graph
under consideration. One may assume that $G$ is free of self-loops,
which do not affect the ordering scores.  Graph $G$ is identified by
its adjacency matrix $A$.  Let $n=|V|$. Let $m={\rm nnz}(A)$, the
number of nonzeros in $A$.  Then, $m=2\,|E|$ when $G$ is undirected
and $m=|E|$ when $G$ is directed.  Denote by $\Pi(n)$ the set of all
permutations of $(1:n)$.  Chierichetti, Kumar and Lattanzi introduced
in 2009~\cite{chierichetti2009compressingsocial} the following two
measures of adjacency access locality (\aal{}) captured by a vertex
ordering $\pi \in \Pi(n)$,
\begin{equation}
  \label{eq:mLogA}
  \small 
 \begin{aligned} 
   \mloga(G, \pi ) & = \frac{1}{m}
    \sum_{(u,v) \in E} \log_2 \left( 1 \!+\!
     \left| \pi(u) - \pi(v) \right| \right)
   \\
   & = \frac{1}{m} 
   \sum_{v\in V}
  \sum_{\stackrel{u \in \mathcal{N}(v)}{u\neq v}}    
  \!\! \log_2 \left(\, 1 \!+\!
     \left| \pi(u) \!-\! \pi(v) \right|\, \right),
 \end{aligned} 
\end{equation}
\begin{equation}
  \label{eq:mLogGapA}
  \small 
  \noindent
  \mloggapa(G, \pi)
  = \frac{1}{m} \sum_{v \in V}
  1 \!+\!\! \sum_{\stackrel{u_i \in \mathcal{N}[v]}{i=2:d(v)}}
  \!\!\! \log_2 \left( 1 \!+\! 
    \pi(u_i) \!-\! \pi(u_{i\!-\!1})  \right), 
\end{equation}
where the neighbors $u_i$ of $v$ are ordered by $\pi$,
$\pi(u_i)>\pi(u_{i-1})$.
We present the second expression in (\ref{eq:mLogA}) in order to make
the adjacency lists more explicit and to make the connection to and
difference from (\ref{eq:mLogGapA}) more salient. We make a slight
modification in (\ref{eq:mLogGapA}) by replacing
$\log_{2}(1+|\pi(u_1)-\pi(v)|)$ with $1$, effectively removing
the host vertex $v$ from its own adjacency list.  This change makes
the measure more closely related to general-purpose compression
schemes, as shown in Figure~\ref{fig:mLogGapA-hdf5-correlation}.
The measure \mloga{} is the average distance, in bit length, of the
neighbors from each host vertex $v$; \mloggapa{} is the average gap in
bit length between the successive neighbors of each host $v$. The goal
is to find the vertex ordering that minimizes one of the measures or
both. However, locating the optimum on an arbitrary graph is
NP-hard~\cite{chierichetti2009compressingsocial} and computationally
intractable. Practical ordering schemes resort to various heuristics.

The average gap measure (\ref{eq:mLogGapA}) is closely related to
subsequent compression schemes. We make a novel use of the average
distance measure (\ref{eq:mLogA}) as well.  The following two basic
inequalities hold for any connected graph $G$,
\begin{equation}
  \label{eq:basic-innequalities} 
1 
  \leq
  \mloggapa(G, \pi) \leq \mloga(G,\pi),
  \,\,\,
  \forall \, \pi \in \Pi(n).
\end{equation}
The absolute difference between the two measures is 
\begin{equation}
  \label{eq:Delta-G-pi} 
  \Delta(G,\pi)\! \triangleq \!\mloga(G,\pi)-\mloggapa(G,\pi).
\end{equation}
We will demonstrate shortly how we utilize the differential
information.  We also use the basic statistics of the graph
degrees. Frequently, we use the average degree $\bar{d}$.
Conventionally, a graph is regarded as sparse if
$\bar{d} \in O(\log_2n)$.

\begin{figure}
  \centering
  \begin{subfigure}{0.20\linewidth}
    \includegraphics[width=\linewidth]
    {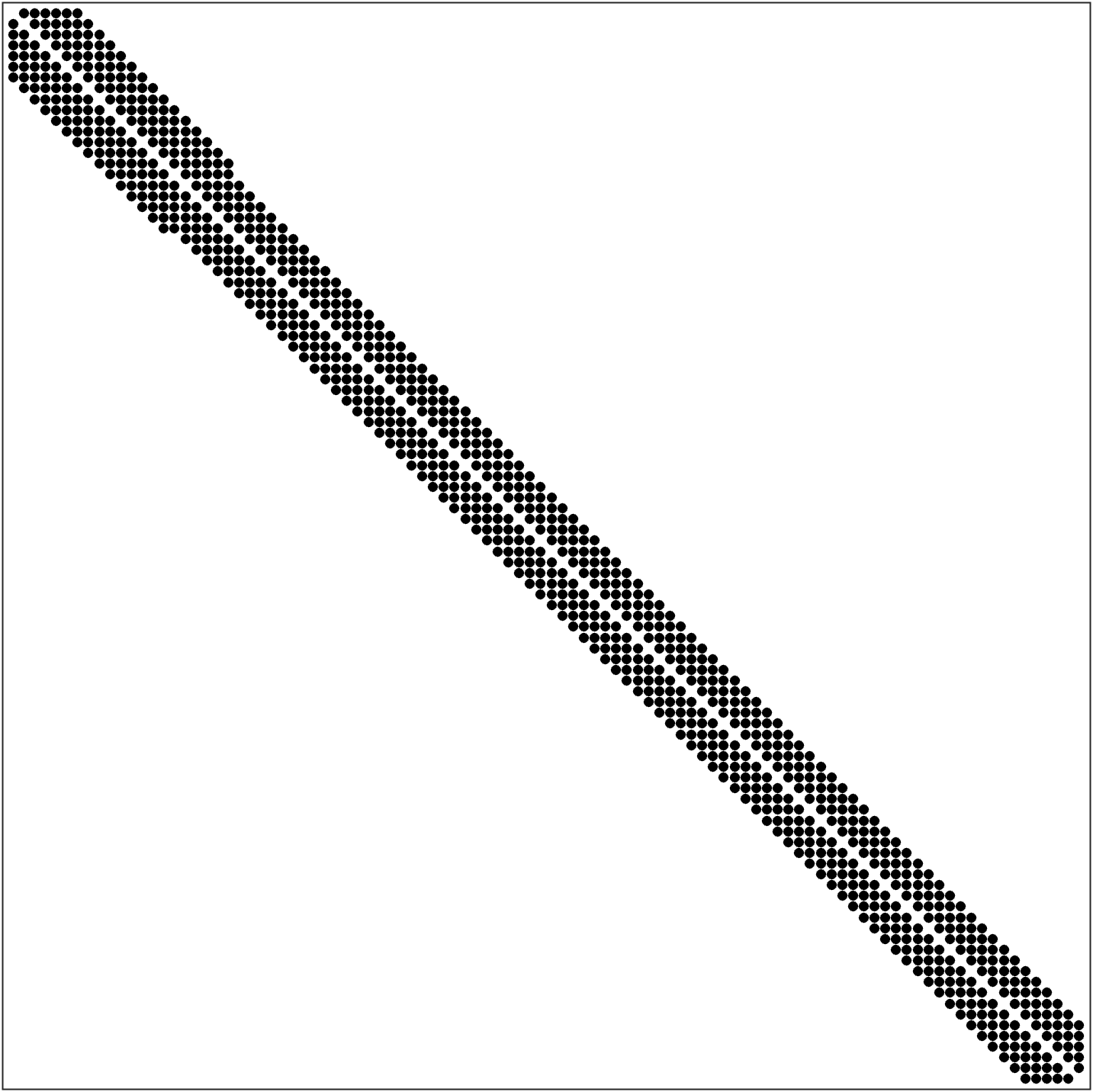}
    \subcaption{$A_{\rm conv1}$}
  \end{subfigure}
  \hspace{.5em}
\begin{subfigure}{0.20\linewidth}
    \includegraphics[width=\linewidth]
    {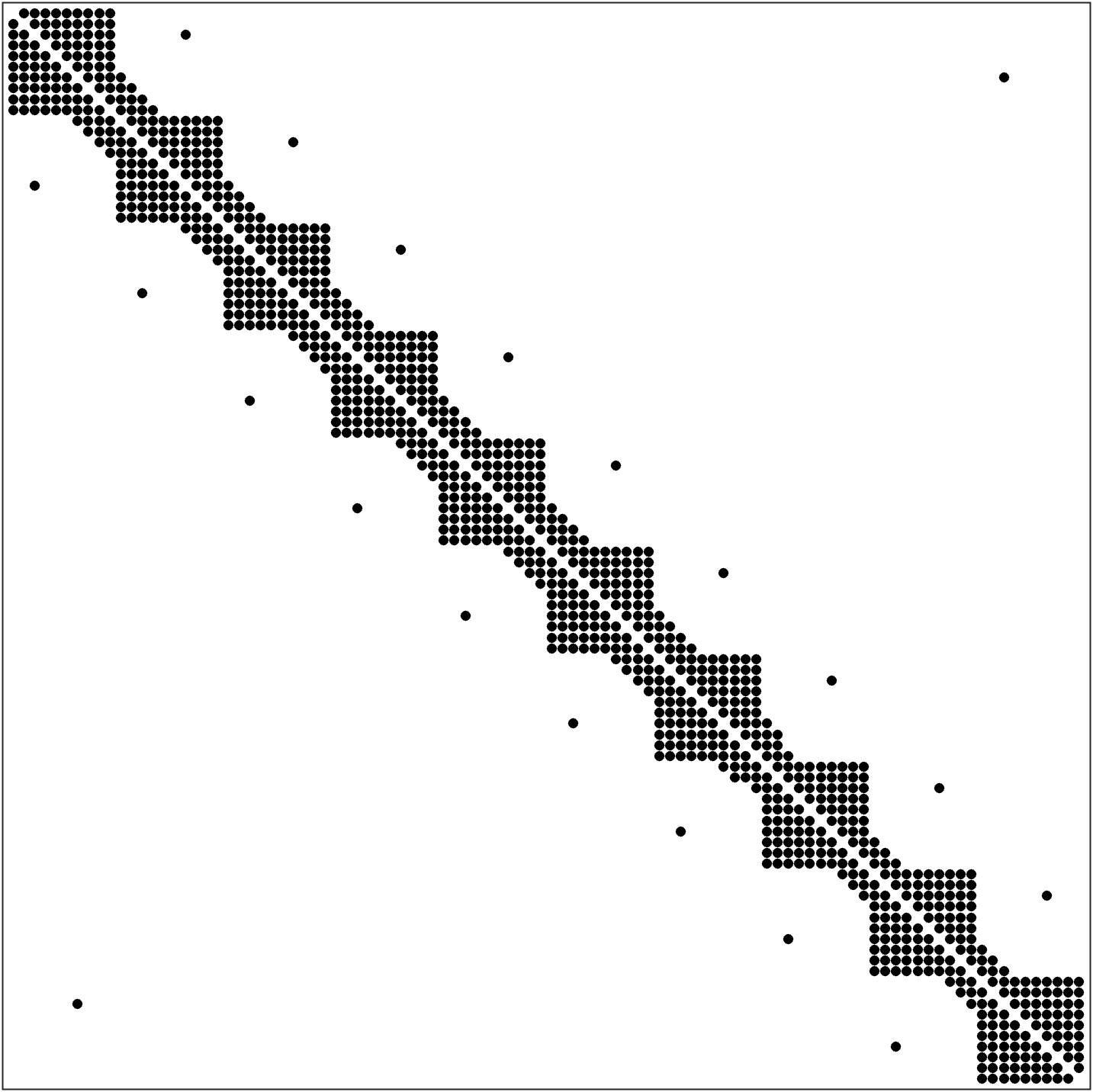}
    \subcaption{$A_{\rm PoK}$} 
  \end{subfigure}
  \hspace{.5em}
\begin{subfigure}{0.20\linewidth}
    \includegraphics[width=\linewidth]
    {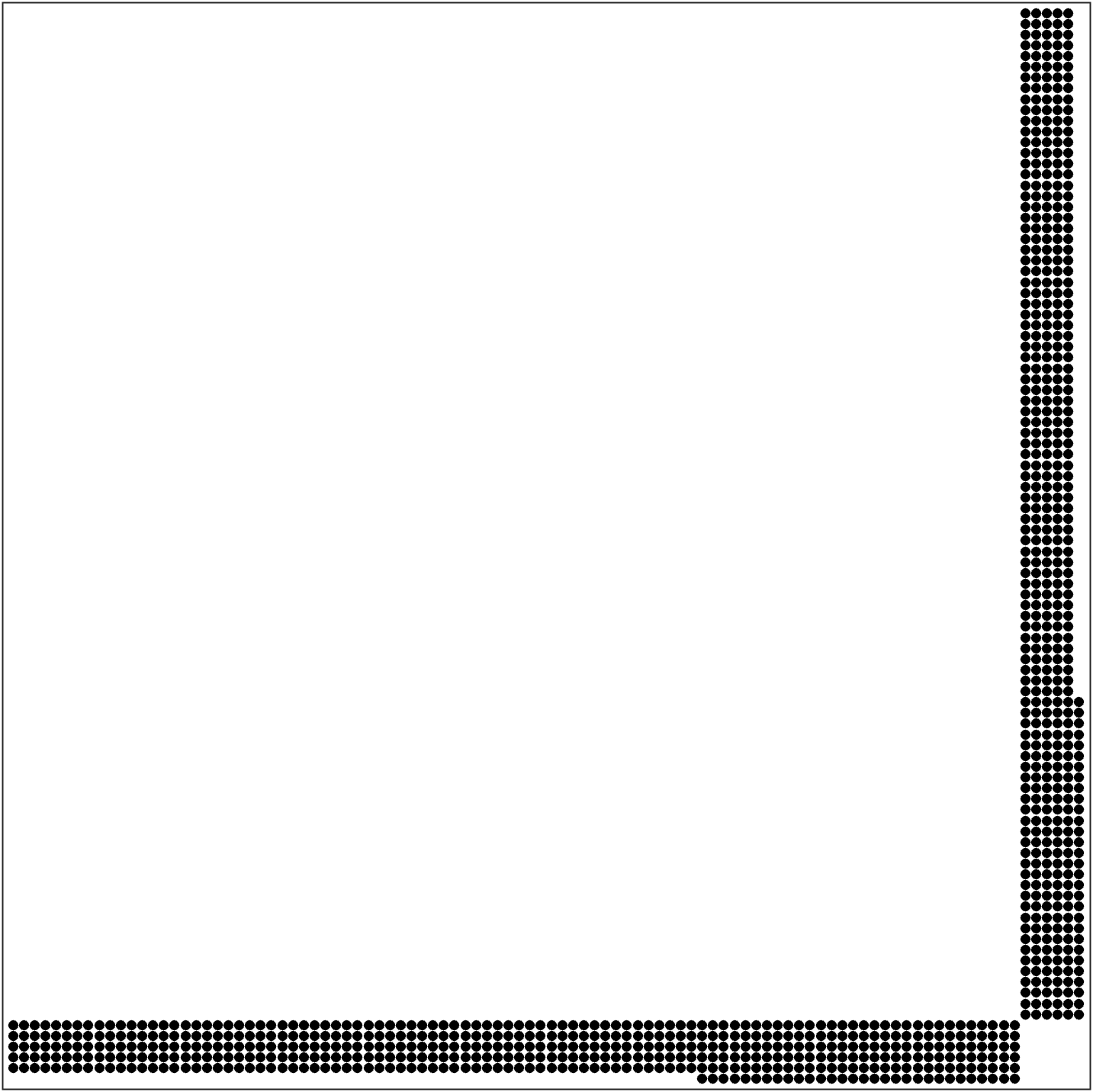}
    \subcaption{$A_{K_{(b,n\!-\! b)}}$}
  \end{subfigure}
  \hspace{.5em}
\begin{subfigure}{0.20\linewidth}
    \includegraphics[width=\linewidth]
    {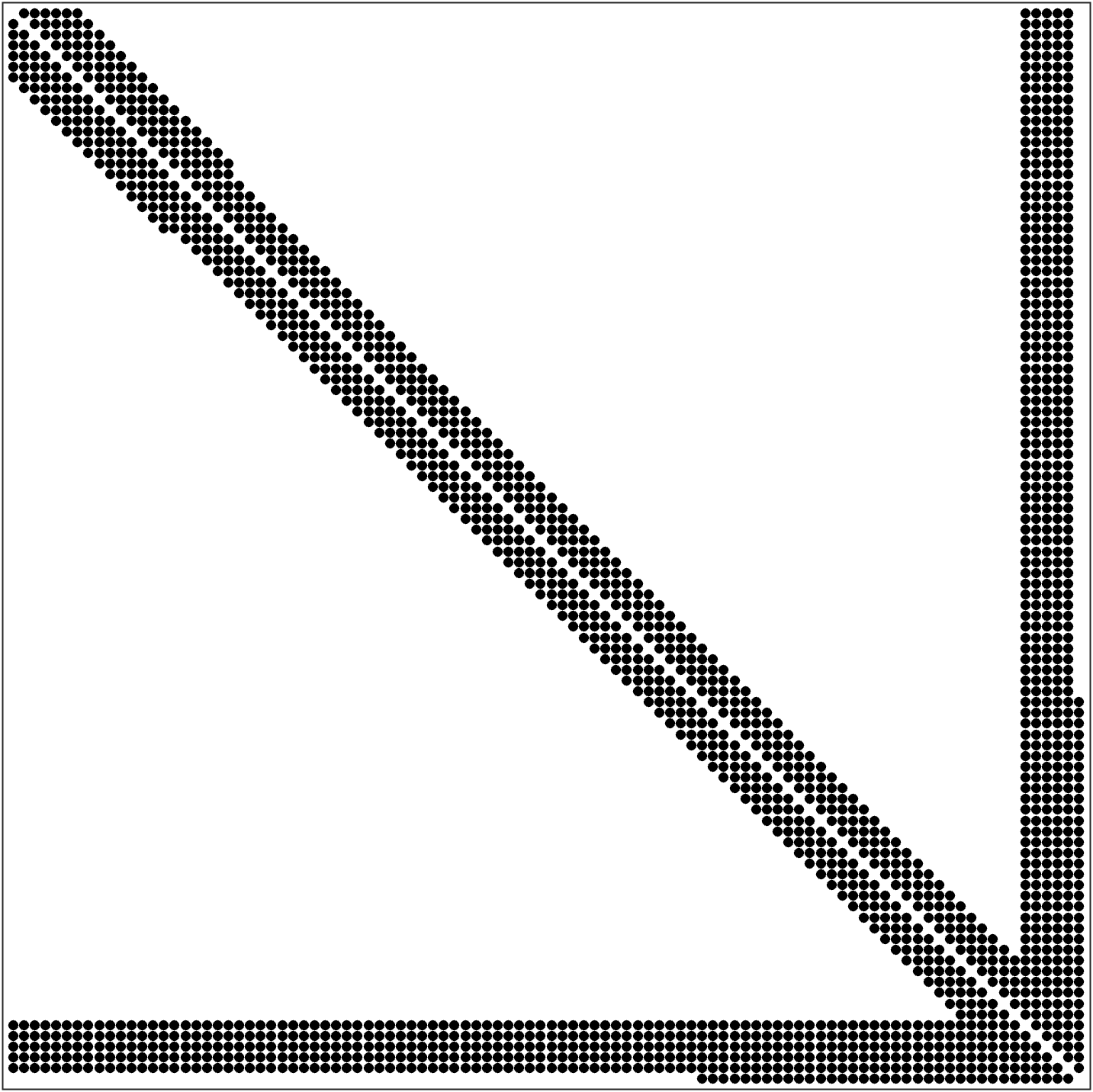}
    \subcaption{$A_{\rm wheel}$} 
 \end{subfigure}
\\ \vskip 0.5em
  \begin{subfigure}[c]{0.31\linewidth}
    \centering
    \includegraphics[width=0.8\linewidth]
    {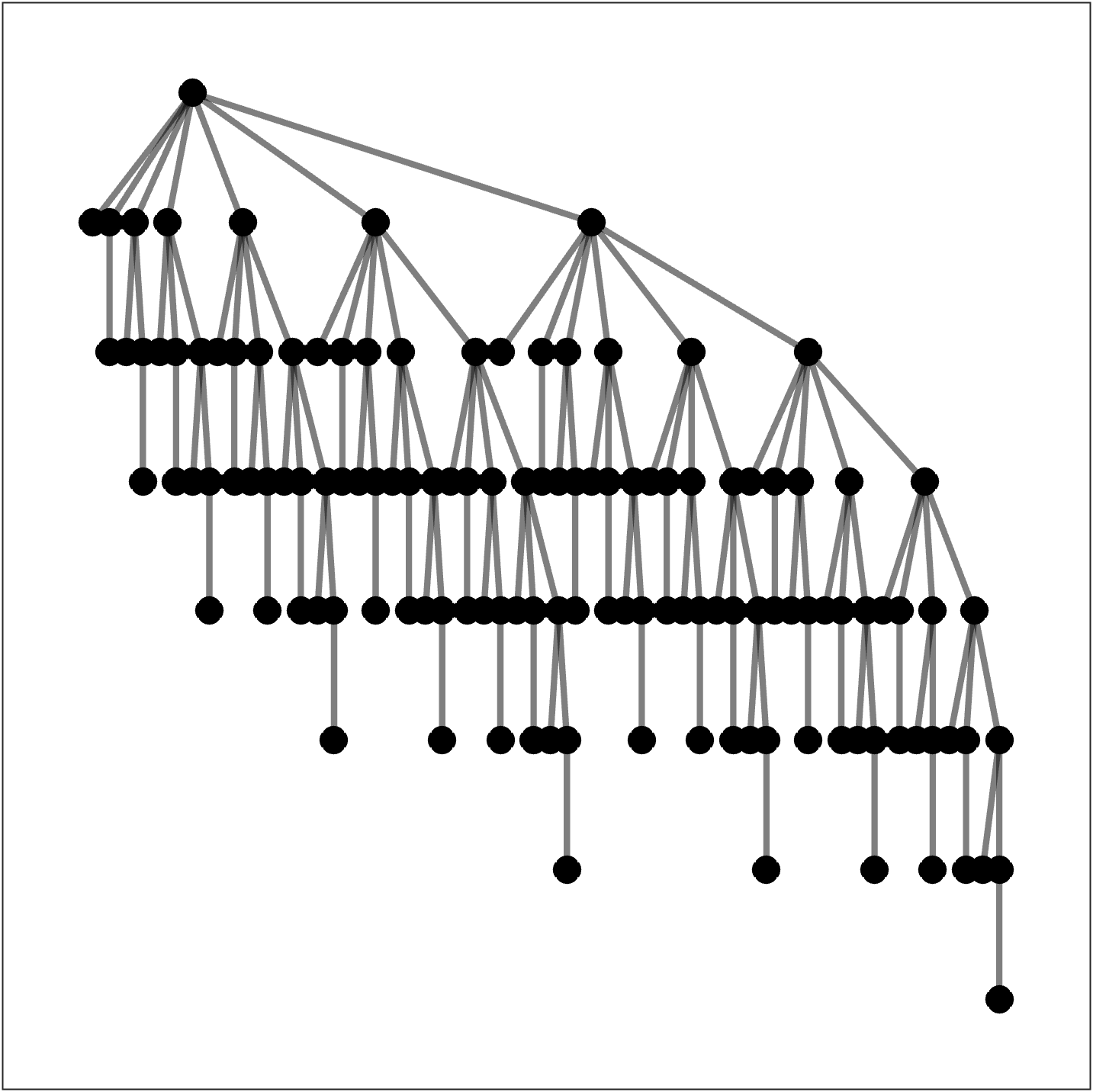}
    \subcaption{$T_{\rm binomial}$}
  \end{subfigure}
\begin{subfigure}[c]{0.31\linewidth}
    \centering
    \includegraphics[width=0.8\linewidth]
    {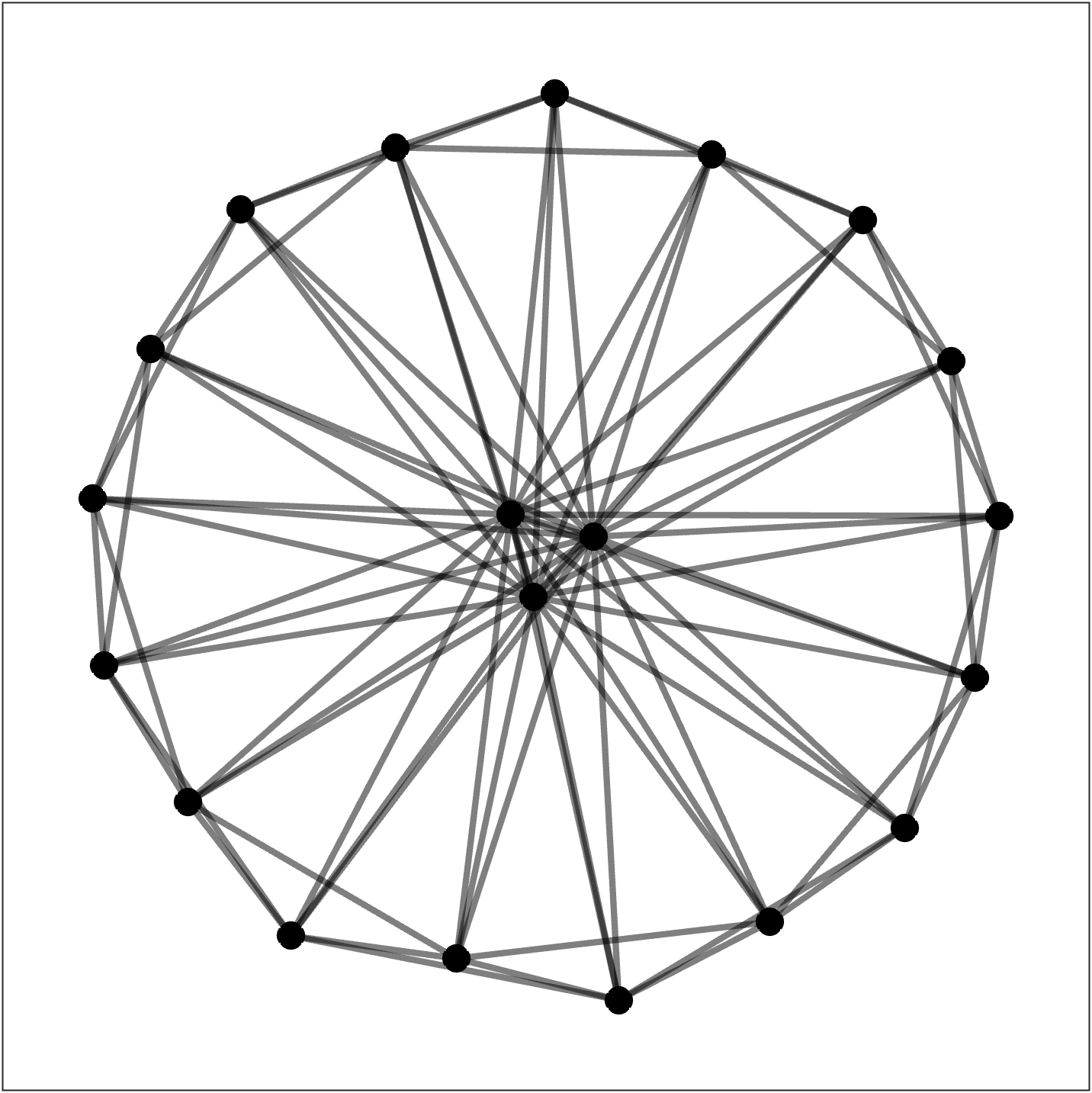}
    \subcaption{$G_{\rm wheel}$}
  \end{subfigure}
\begin{subfigure}[c]{0.31\linewidth}
    \centering
    \includegraphics[width=0.9\linewidth]
    {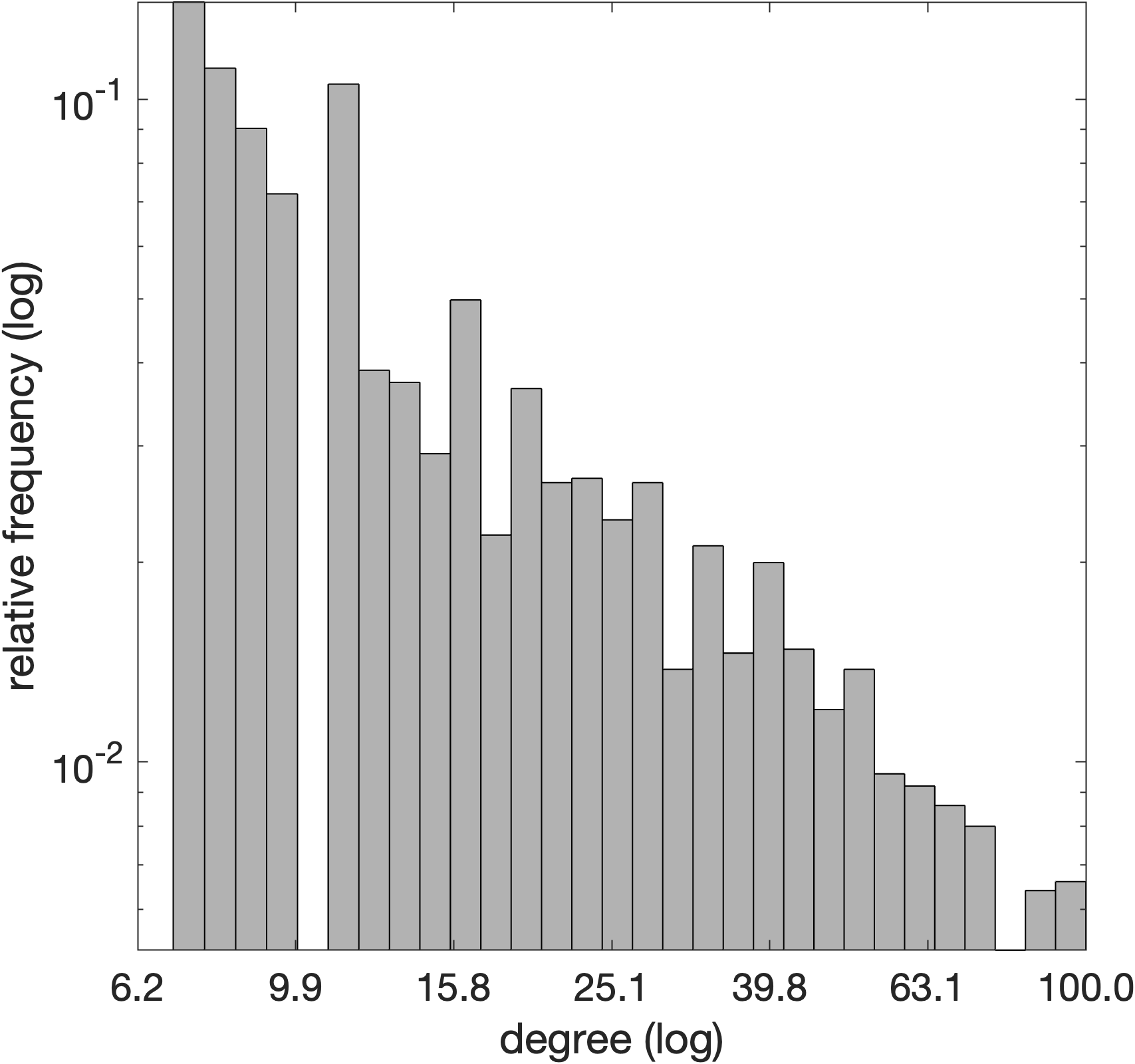}
    \subcaption{$d(G_{\rm lfr})$}
  \end{subfigure}
\vspace{-0.5em}
\caption{\text{\bf Top:} The adjacency matrices of $G_{\rm conv1}$,
    $G_{\rm PoK}$, $K_{(b,n\!-\!b)}$ and $G_{\rm wheel}$ in their
    respective compact forms. \text{\bf bottom:} $T_{\rm binomial}$ and $G_{\rm wheel}$ in 2D
    spatial displays, and the degree distribution of
    graph $G_{\rm LFR}$ in log-log scale. } \label{fig:depiction-benchmark-graphs} 
  \vspace{-1.5em}
\end{figure}

\subsection{Elementary compact sparse graphs}
\label{sec:elementary-compact-sparse-graphs}

We make a formal \aal{} analysis of vertex ordering from the
perspective of graph composition or decomposition.  This perspective
is not foreign.  Beneath their differences in algorithms and results,
existing vertex ordering methods are common in that each has a
strategy for extracting certain subgraphs that are more compressible,
and it is used in tandem with an ordering priority that such subgraphs
be overlapped as much as possible. We make an explicit \aal{}
analysis of three elementary types of sparse graphs that are
frequently used to prototype and approximate subgraphs in a larger
graph and are highly compressible.  The corresponding adjacency
matrices in their respective optimal vertex orderings are among the
most compact sparse matrices.

First, the convolution network/graph with a narrowly banded adjacency
matrix is perhaps a more familiar type for subgraphs of a large sparse
graph.  When $\bar{d}>2$, the one-dimensional convolution graph
$G_{\rm conv1}$ is next to the highest sum of local cluster
coefficients. It is the base reference for modeling and generating
(via edge rewiring) the small-world networks $G_{\rm ws}$ by the
Watts-Strogatz model~\cite{watts1998collectivedynamics}.  The
adjacency matrix $A$ in the \rcm{} ordering $\pi_{\rcm}$ is banded
with the minimal bandwidth.  If $G_{\rm conv1}$ is not
circulant, then
\begin{equation}
  \label{eq:conv1-mLogGapA-at-rcm}
  \small 
  \begin{aligned}   
    & \mloggapa(G_{\rm conv1}, \pi_{\rcm})
     =
    \min_{\pi\in \Pi(n)} \mloggapa(G_{\rm conv1},\pi),
    \\
    &
    \phantom{xxxxxxxxxxxxxxxxxxx}
    =     1 + \gamma_{1} \dfrac{\log_2(3)-1}{\bar{d}}
  \end{aligned}     
\end{equation}
and 
\begin{equation}
  \label{eq:conv1-mLogA-at-rcm}
  \small 
  \begin{aligned}   
    & \mloga(G_{\rm conv1}, \pi_{\rcm}) =
    \min_{\stackrel{G:m/n=\bar{d}}{\pi\in \Pi(n)}}
      \mloga(G,\pi),
    \\
    & \frac{b-1}{b} \log_2\left(\frac{b}{2}\right)
    \leq
    \Delta (G_{\rm conv1}, \pi_{\rcm})
    \leq \log_2(b),
  \end{aligned}     
\end{equation}
where $b$ is the semi-bandwidth, $b = \lceil\bar{d}/2\rceil$.  Here,
$\gamma_{1}$ denotes a value close to $1$ and varies little with
$\bar{d}$.

Next is the graph $G_{\rm PoK}(n,\bar{d})$ constructed as a linear
path/chain of cliques $K_{\bar{d}}$, see
Figure~\ref{fig:depiction-benchmark-graphs}. It is a prototype sparse graph
consisting of equally populated sub-communities with the maximal
intra-community connection and the minimal inter-community
connection~\cite{fortunato2007resolution,traag2011}. The removal of
any inter-clique link decouples the graph.  The adjacency matrix
$A_{\rm PoK}$ is of nearly block diagonal form by the path ordering,
which is also of the minimal bandwidth.  Although the semi-bandwidth
is $\bar{d}$ instead of $\bar{d}/2$, the number of bits per link is
doubled.  The difference from the minimal on $G_{\rm conv1}$ is small
in \mloga{}, no greater than $\log_2(b)/\bar{d}$ bits per link, and
even smaller in \mloggapa, no greater than
$\log_2(b)/(1+\bar{d}^2/2)$.  In short, $G_{\rm PoK}$ can be treated
almost the same as $G_{\rm conv1}$.

Thirdly, the most primitive for modeling compact subgraphs of a
sparse graph/network is the biclique $K_{(b,n\!-\! b)}$.  In the
special case $b=1$, $K_{(1,n\!-\!1)}$ is the star graph. Any tree
graph is composed of star subgraphs. For $b>1$, the biclique
$K_{(b,n-b)}$ can be viewed as a block version of a star graph with
$b$ center nodes, every center node is connected to all $n\!-\!b$
peripheral nodes.  If the sparsity is specified by $\bar{d}$, the
biclique is assumed to have $m=\bar{d}n$ links with
$b=\lceil \bar{d}/2 \rceil$, it is incomplete if $\bar{d}$ is not an
even integer.  In the case of $\bar{d} = O(\log n)$, $K_{(b,n\!-\!b)}$
has a large gap in the degrees between the $b$ center nodes and the
rest.  This degree inequality is a typical phenomenon in many social
or biomolecular networks.  The adjacency matrix of $K_{(b,n\!-\!b)}$
is of low rank, regardless of the ordering, while its counterparts for
$G_{\rm conv1}$ and $G_{\rm PoK}$ are nearly regular and close to full
rank. In its compact form, the adjacency matrix has a block row and a
block column when the center nodes are indexed together, such as by
$\pi_{\amd}$, the \amd{} ordering. There are several remarkable
properties with the biclique compression.
\begin{proposition}
  \label{prop:min-mLogGapA-overall}
  Among all sparse graphs with the average degree $\bar{d}$ and over
  $\Pi(n)$ of any size $n$, the minimal \mloggapa{} score is
  achieved by $\pi_{\amd}$ on $K_{(b,n\!-\!b)}$,
  $\!\small b\!=\!\left\lceil\bar{d}/2\right\rceil\!$,
\begin{equation}
  \label{eq:mLogGapA-lower-bound}
  \small
  \min_{\stackrel{G:m/n\!=\!\bar{d}}{\pi\in \Pi(n)}}\!\! 
   \mloggapa(G,\pi) 
   \!=\!\mloggapa(K_{(b,n-b)}, \pi_{\amd{}}) \!=\! 1.    
\end{equation}
Moreover, there is the minimum-equivalence set that admits any
ordering $\pi$ on $K_{(b,n\!-\!b)}$ if, and only if, $\pi$ places the
$b$ center nodes together.
\end{proposition}
By the proposition, any non-circulant shift of $\pi_{\amd}$ is an
optimal ordering on $K_{(b,n\!-\!b)}$.  Regarding the necessary
condition, if the high-degree nodes are dispersed far apart by an
ordering $\pi$, the score can be as large as, although bounded by,
$\mloga(K_{(b.n\!-\!b)}, \pi_{\amd})$.  We omit the special detail and
present the general relations, beyond (\ref{eq:basic-innequalities}),
between the two scores by \mloga{} and \mloggapa.
\begin{proposition}
  \label{prop:max-mLogGapA-overall} 
  The feasible \mloggapa{} scores on all sparse graphs of the same
  average degree $\bar{d}$ and over $\Pi(n)$ of any size $n$ are
  closely bounded from above as follows, $b=\lceil\bar{d}/2\rceil$,
\begin{equation}
  \label{eq:mLogGapA-upper-bound}
  \small
  \begin{array}{rl}
    \displaystyle 
  \max_{\stackrel{G:m/n\!=\!\bar{d}}{\pi\in \Pi(n)}}\!\! 
    \mloggapa(G,\pi)\! 
    & \leq \mloga(K_{(b,n-b)}, \pi_{\amd{}})
    \\
    & \displaystyle 
      = 1 + \gamma_{1} \log_2(1\!+\!n\!-\!b).
\end{array} 
\end{equation}
\end{proposition} 

\begin{proposition}
  For an ordering $\pi$ on graph $G$, the $2$-tuple
  $[\,\mloggapa(G,\pi),\Delta(G,\pi) \, ]$ is a descriptor of
  the substructures of $A(\pi,\pi)$, a larger relative difference
  indicates a larger portion of non-zero elements being off from the
  $\bar{d}(G)$ main diagonals.
\end{proposition}
In more detail, when \mloggapa{} is close to the baseline at
$1$, then $A(\pi,\pi)$ is nearly banded or block-diagonal. If
$\Delta (G,\pi)$ is close to $0$ in addition, then $A(\pi,\pi)$ is
narrowly banded or block-diagonal within a narrow band. At the other
extreme, the following condition,
\begin{equation}
  \label{eq:warning-condition}
  \mloga(G,\pi) >
  \mloga(K_{(b,n\!-\!b)},\pi_{\amd})
\end{equation}
suggests an improvement or replacement of the ordering $\pi$.

\begin{table*}[]
  \centering
  \caption{Empirical assessment of adjacency access localities and subgraph
    structures captured by \num{6} different vertex ordering schemes
    on \num{18} sparse graphs of diverse types.  \slashburn{} is the baseline. Each graph $G$ has $|V|$ vertices and average degree $\bar{d}$.
    The number of nonzero elements in the adjacency matrix $A$ is
    $\text{nnz}(A) = \bar{d}|V|$. For the synthetically generated
    graphs, except the binomial tree $G_{\rm binomial}$, $|V|=250K$
    with $\si{K}=\num{1000}$ and $\bar{d}=14$.
Each table entry is a $2$-tuple descriptor
    $[\, \mloggapa(G,\pi) \mid \Delta(G,\pi) \, ]$. 
    The first element is the number of bits per link on average
    by ordering $\pi$ on graph $G$, the sum of the two elements
    is the \mloga{} score, namely,  the average neighbor
    distance in bit length, and the ratio of the first element
    to the second indicates the average portion of links  within
    the $\bar{d}/2$ range in each adjacency list, as stated in
    Proposition~\ref{prop:max-mLogGapA-overall}.
    On each graph, the best \mloggapa{} score is highlighted
    in bold case, the runner-up is underlined.
Figure~\ref{fig:aal-measures} depicts the table entries for three of the graphs. 
\label{tab:schemes-orders}
  }
\footnotesize
\begin{tabular}{
    @{}
    l
    @{}
    l
    @{\,\,}
    r
    @{\,\,}
    S[tight-spacing=true,table-alignment-mode=none,
    table-number-alignment=right,
    table-text-alignment=right,
    drop-zero-decimal=true,
    ]
    @{\,}
    l
    S[tight-spacing=true,table-alignment-mode=none,
    table-number-alignment=right,
    table-text-alignment=right,
    drop-zero-decimal=true,
    ]
    @{\,}
    l
    S[table-format=2.0,round-mode=places,round-precision=0,mode=text]
    @{\quad\quad\quad}
S[table-format=1.1,round-mode=places,round-precision=1,mode=text,detect-weight=true,detect-inline-weight=math,text-series-to-math]
    @{\makebox[7pt]{$|$}}
    S[table-format=2.1,round-mode=places,round-precision=1,mode=text]
S[table-format=1.1,round-mode=places,round-precision=1,mode=text,detect-weight=true,detect-inline-weight=math,text-series-to-math]
    @{\makebox[7pt]{$|$}}
    S[table-format=2.1,round-mode=places,round-precision=1,mode=text]
S[table-format=1.1,round-mode=places,round-precision=1,mode=text,detect-weight=true,detect-inline-weight=math,text-series-to-math]
    @{\makebox[7pt]{$|$}}
    S[table-format=2.1,round-mode=places,round-precision=1,mode=text]
S[table-format=1.1,round-mode=places,round-precision=1,mode=text,detect-weight=true,detect-inline-weight=math,text-series-to-math]
    @{\makebox[7pt]{$|$}}
    S[table-format=2.1,round-mode=places,round-precision=1,mode=text]
S[table-format=1.1,round-mode=places,round-precision=1,mode=text,detect-weight=true,detect-inline-weight=math,text-series-to-math]
    @{\makebox[7pt]{$|$}}
    S[table-format=2.1,round-mode=places,round-precision=1,mode=text]
@{\quad\quad\quad}
    S[table-format=1.1,round-mode=places,round-precision=1,mode=text,detect-weight=true,detect-inline-weight=math,text-series-to-math]
    @{\makebox[7pt]{$|$}}
    S[table-format=2.1,round-mode=places,round-precision=1,mode=text]
    @{}
    }
    \toprule
    & &
    & \multicolumn{2}{c}{$n\!=\! |V|$} & \multicolumn{2}{c}{$\textrm{nnz}(A)$}
    & {$\bar{d}$}
    & \multicolumn{2}{c}{\rcm\quad\quad} 
    & \multicolumn{2}{c}{\fcutone} 
    & \multicolumn{2}{@{\hspace*{0.0em}}c}{\sf SlashBurn}
    & \multicolumn{2}{c}{\ned} 
    & \multicolumn{2}{c}{\amd\quad\quad\quad}  
    & \multicolumn{2}{c}{\vifps} 
    \\ 
    \midrule
\parbox[t]{6mm}{\multirow{7}{*}{\rotatebox[origin=c]{90}{synthetic}}}
  & $G_{\rm conv1}$                                    &
  & 250                                                & \si{K}   & 3.5   & \si{M}  & 14       & \UU{1.0}   & 1.14   & 3.25       & 1.75   & 11.36  & 4.00   &  1.65      & 1.61   & \UU{1.04}  & 1.14   & \B 1.04    & 1.14
    \\
  & $G_{\rm PoK}$                                      &
  & 250                                                & \si{K}   & 3.5   & \si{M}  & 14       & \UU{1.06}  & 1.30   & 5.11       & 4.67   & 10.51  & 3.70   &  1.67      & 2.29   & 1.09       & 1.33   & \B 1.06    & 1.30
    \\
  & $K_{(7,n-7)}$                                      &
  & 250                                                & \si{K}   & 3.5   & \si{M}  & 14       & 1.00       & 15.55  & 1.00       & 15.55  & 1.00   & 15.56  & 1.00       & 15.56  & 1.00       & 15.56  & 1.00       & 15.56
    \\
  & $G_{\rm wheel}$                                    &
  & 250                                                & \si{K}   & 3.5   & \si{M}  & 14       & 3.11       & 13.28  & 3.08       & 13.31  & 3.11   & 13.27  &  2.46      & 12.38  & \UU{2.15}  & 12.13  & \B 2.15    & 12.13
    \\
  & $G_{\rm ws}$                                       &
  & 250                                                & \si{K}   & 3.5   & \si{M}  & 14       & 5.30       & 5.33   & 9.64       & 2.76   & 11.41  & 4.34   &  5.30      & 3.54   & \UU{5.01}  & 3.86   & \B 2.54    & 2.34
    \\
  & $G_{\rm lfr}$                                      &
  & 250                                                & \si{K}   & 3.5   & \si{M}  & 14       & \UU{5.38}  & 8.83   & 6.31       & 8.19   & 6.79   & 8.25   &  7.03      & 8.20   & 6.28       & 8.56   & \B 5.09    & 4.74
    \\
  & $T_{\rm binomial}$                                 &
  & 262                                                & \si{K}   & 0.52  & \si{M}  & 2        & 4.57       & 9.69   & 7.59       & 6.59   & 8.06   & 7.76   &  2.77      & 1.69   & \UU{1.86}  & 0.94   & \B 1.65    & 0.57
    \\
\midrule[.2pt]
    \parbox[t]{6mm}{\multirow{11}{*}{\rotatebox[origin=c]{90}{real-world}}}
  & Polbooks~\cite{newman2006modularitycommunity}      &
  & 105                                                & \si{}    & 882   &         & 8.4      & \B 1.84    & 1.44   & 1.94       & 1.72   & 2.33   & 2.29   & 1.96       & 1.88   & 2.00       & 2.08   & \UU{1.87}  & 1.22
    \\
  & Football~\cite{girvan2002communitystructure}       &
  & 115                                                & \si{}    & 613   &         & 5.33     & 2.26       & 1.70   & \UU{2.21}  & 1.48   & 2.61   & 2.49   & 2.45       & 1.99   & 2.35       & 2.16   & \B 1.93    & 1.11
    \\ 
  & URV email~\cite{guimera2003selfsimilarcommunitya}  &
  & 1.1                                                & \si{K}   & 10.9  & \si{K}  & 9.622    & 4.00       & 3.07   & 3.98       & 2.46   & 4.30   & 3.18   & \UU{3.78}  & 2.96   & \UU{3.77}  & 2.92   & \B 3.27    & 1.89
    \\
  & Polblogs~\cite{adamic2005politicalblogosphere}     &
  & 1.2                                                & \si{K}   & 19.0  & \si{K}  & 15.565   & \UU{.97}   & 5.94   & 3.24       & 4.01   & 3.38   & 5.26   & 3.45       & 5.14   & 2.94       & 5.63   & \B .97     & 5.94
    \\
  & Powergrid~\cite{watts1998collectivedynamics}       &
  & 4.9                                                & \si{K}   & 13.2  & \si{K}  & 2.669    & 3.48       & 3.28   & 4.03       & 2.98   & 4.14   & 2.84   & 2.94       & 2.02   & \UU{2.34}  & 1.04   & \B 2.07    & 0.46
    \\
  & Pothen-Barth~\cite{pothen1997graphpartitioning}    &
  & 6.7                                                & \si{K}   & 46.2  & \si{K}  & 6.90285  & 2.91       & 1.34   & 4.79       & 1.08   & 6.79   & 1.90   & 3.13       & 1.07   & \UU{2.49}  & 1.08   & \B 2.06    & 0.72
    \\
  & Enron email~\cite{leskovec2005graphstime}          &
  & 36                                                 & \si{K}   & 361   & \si{K}  & 10.6176  & 4.84       & 5.72   & 5.46       & 5.33   & 5.69   & 5.27   & 4.73       & 5.43   & \UU{4.52}  & 5.37   & \B 3.93    & 5.48
    \\
  & APS-2020~\cite{aps-2020}                            &
  & 0.67                                               & \si{M}   & 8.85  & \si{M}  & 13.247   &  8.01      & 8.45   & 11.21      & 5.23   & 10.98  & 5.90   & \UU{6.88}  & 8.60   & 8.27       & 8.55   & \B 5.63    & 10.19
    \\
  & Flickr~\cite{gleich2012}                           &
  & 0.82                                               & \si{M}   & 9.84  & \si{M}  & 11.9817  & \UU{5.97}  & 11.75  & 6.45       & 7.21   & 6.84   & 7.10   & 6.24       & 10.77  & 6.12       & 11.69  & \B 5.29    & 12.24
    \\
    & YouTube~\cite{yang2015Definingevaluating}        &
  & 1.1                                                &  \si{M}  & 6     & \si{M}  &  5.27    & 7.48       & 9.11   & 7.83       & 8.37   & 8.59   & 7.77   & 8.01       & 6.63   & \UU{7.26}       & 6.85   & \B 7.26       & 6.85
    \\
    & LiveJournal~\cite{yang2015Definingevaluating}    &
  & 4                                                  &  \si{M}  & 69.4  & \si{M}  & 17.35    & 9.82       & 7.30   & 12.17      & 5.55   & 12.73  & 5.48   & \UU{9.10}       & 5.90   & 10.77      & 5.83   & \B 9.10       & 5.90
    \\
  \bottomrule
  \end{tabular}
\vspace*{-2em}
\end{table*}

\subsection{Composition effects}

Among numerous ways of graph composition, we make a simple abstraction
of their effects on the adjacency locality in vertex ordering. Fix the
size $n=|V|$.  Let graph $G_{\rm wheel}(n,b_{l},b_{g} ) $ be the sum
of $K_{(b_{g},n\!-\!b_{g}) }$ and $G_{\rm conv1}(b_{l})$ of
semi-bandwidth $b_{l}$, $b_{l}b_{g}>0$.  Then,
$\bar{d}(G_{\rm wheel}) = 2(b_{l}\!+\!b_{g})$. Among the $n-b_{g}$
low-degree vertices, each has $2b_{l}$ neighbors that are locally
connected and has direct links to the $b_{g}$ global center nodes. The
$b_{g}$ center nodes have nearly half or more of the total links
when $b_{g} \geq b_{l}$.  Graph $G_{\rm wheel}$ is of a wheel shape,
see Figure~\ref{fig:depiction-benchmark-graphs}. In terms of local and
global connectivities, $G_{\rm wheel}$ combines the best features from
each component graph. In fact, it is the small-world graph with the
maximal sum of the local cluster coefficients and the minimal diameter
among all graphs of the same size and sparsity. In the context of
graph compression, however, the wheel composition effect may be both
startling and elucidating.  The lowest gap score on $G_{\rm wheel}$
can be much greater than the sum of the individual components,
\begin{equation}
  \label{eq:wheel-mLogGapA-min}
  \small 
  \mloggapa(G_{\rm wheel}, \pi_{\amd{}})
  = 1+ \gamma_{1} \frac{\log_2(n\!-\!\bar{d})}{\bar{d}}, 
  \quad b_1b_2 > 0.  
\end{equation}
The least number of bits per link for $G_{\rm wheel}$ increases with
$\log_2(n)/\bar{d}$ roughly, whereas it is $1$ for
$K_{(b_{g},n\!-\!b_{g})}$ and close to $1$ for $G_{\rm conv1}(b_{l})$,
invariant to variation in $n$.  Due to the sequential nature of vertex
ordering, the composition inevitably introduces gaps in every
adjacency list between the $b_{l}$ locally-connected grass-roots
neighbors and the $b_{g}$ global center nodes. The gap length varies
from $1$ up to $n\!-\!\bar{d}/2$.

The composition effect expressed by (\ref{eq:wheel-mLogGapA-min}) also
has brighter implications. First, if $G_{\rm wheel}$ is not super
sparse, such as when $\bar{d} \geq \log_{2}(n)$, then it is highly
compressible.  Secondly, a large and sparser network in the real world
is typically composed of smaller subnetworks/subgraphs across multiple
layers/levels, the grass-roots nodes do not necessarily have direct
links to the elite nodes at the top. Consider the wheel graphs as
subgraphs of a larger network.  For instance, let $G$ be a chain of
wheel-shaped subgraphs of size $n_s$.  Then, the number of extra bits
per link is roughly $\log_2(n_s)/\bar{d}$ instead of
$\log_2(n)/\bar{d}$. Here, the assumed homogeneity in the subgraph
size and shape is for brevity in describing the
composition/decomposition effects.

Our abstraction of graph composition and decomposition with regard to
vertex ordering for graph compression helps algorithmic thinking or
rethinking.  A Delaunay triangulation graph can be viewed as a
composition of wheel sub-graphs and star subgraphs, although not
necessarily in a single chain. In general, a large, sparse graph can be
decomposed into elementary subgraphs and wheel-like graphs, which may
have missing spokes or missing arcs, the interconnection between the
subgraphs is sparser.

We utilize the information and insight from the above analysis for
evaluating and developing a vertex ordering scheme, as well as for
inferring the local and global connectivity structures of a
graph/network. We use particularly the important reference values in
(\ref{eq:mLogGapA-lower-bound}), (\ref{eq:mLogGapA-upper-bound}),
(\ref{eq:warning-condition}) and (\ref{eq:wheel-mLogGapA-min}).

\section{Performance assessment of existing methods}
\label{sec:existing-limitations}

We report our empirical assessment of five existing vertex ordering
schemes on their effectiveness and specificity/versatility. We use
$\num{18}$ sparse graphs for the assessment.  An ordering scheme is
effective for an intended class of graphs if it maintains decent
performance over variations within the class. A scheme is more
versatile if it is effective across multiple types or classes of
graphs.  In Table~\ref{tab:schemes-orders}, we list the schemes, the
graph types and sizes, data sources, and tabulate the \mloggapa{}
scores and the differential scores. We provide illustrations in
Figure~\ref{fig:aal-measures}.  We describe below the rationale behind
the chosen schemes and graphs and present our findings.

The five vertex ordering schemes are \rcm{}, \fcutone{}, \slashburn{},
\amd{}, and \ned{}, as discussed in Section~\ref{sec:introduction}.  We
denote by \fcutone{} the vertex ordering in the Fiedler eigenvector of
the normalized graph Laplacian.  It is commonly used for graph cut
(binary partition) in parallel sparse matrix
computation~\cite{pothen1990partitioningsparse}.  We introduce the
comparisons with \amd{} and \ned{}, besides \rcm{} and \fcutone{}. The
underlying idea of \amd{} is to approximately minimize the degrees in
successive elimination
graphs~\cite{amestoy2004algorithm837,amestoy1996approximateminimum}.
In the context of vertex ordering for graph compression, the idea can
be interpreted and understood as {\em ordering the vertices from the
  least shared connections/dependencies to the most common
  connections/dependencies.}  \ned{}~(nested dissection) applies
\amd{} to the graph minors contracted by recursive bisections, each
node of a graph minor is a subgraph of the original graph $G$.  The
graph dissection and contraction are proven effective on graphs of
geometrically two/three-dimensional
connectivity~\cite{karypis1998fasthigha}.
The \shingle{} scheme~\cite{chierichetti2009compressingsocial} is not
included as it behaves similarly and is reportedly inferior to
\slashburn{}~\cite{lim2014slashburngraph}.
Unlike the other four schemes, \slashburn{} requires a hyperparameter,
which specifies the number of high-degree nodes to be removed from
every connected component, recursively. We locate the best parameter
value we could for \slashburn{} in every experiment.

The graphs used for the assessment include $\num{7}$ synthetic graphs
for controlled studies and $11$ real-world networks for applicability
assessment, see Table~\ref{tab:schemes-orders}. The first four
synthetic graphs are described in Section~\ref{sec:aal-analysis}.  The
additional three represent, respectively, well-recognized
graph/network types for real-world networks and/or technological
networks.
Graph $G_{\rm WS}$ is a Watts-Strogatz graph with relatively high
local cluster coefficients and a small diameter of $O(\log(n))$.
Graph $G_{\rm LFR}$ is generated by the LFR
simulator~\cite{lancichinetti2008benchmarkgraphsa}, with BA-like
subcommunities of different sizes interconnected by sparser, random
links.
Graph $T_{\rm binomial}$ is a binomial tree typically used in
algorithms/architectures for network routing, priority queuing, and
option pricing, among other
applications~\cite{cormen2022introductionalgorithms}.
The real-world networks are among the most frequently used for
benchmarking, except graph \aps{}.  Covering the \text{APS}
publications over the time span longer than a century, the citation
graph \aps{}, as depicted in
Figure~\ref{fig:teaser-aps2020}, is incredibly valuable in its own
right to network studies.
For each graph, we shuffle its adjacency matrix randomly in order to
reduce the influence of the given or generation sequence on \rcm{},
\amd{}, and \ned{}. The other schemes, \slashburn{} and \fcutone{},
are invariant to the initial ordering.

There are four key takeaways from Table~\ref{tab:schemes-orders}. \begin{inparaenum}[({\sc d}-i)]
\item On the three elementary graphs, all schemes reach the ideal
  score of (\ref{eq:mLogGapA-lower-bound}) on the biclique, the first
  two graphs expose the specificity and limitation of \slashburn{} and
  \fcutone{}, which are far short of reaching the ideal score of
  (\ref{eq:conv1-mLogGapA-at-rcm}).
\item On the small-world graphs, the performance of \slashburn{}
  degrades more from $G_{\rm wheel}$ of diameter $2$ to $G_{\rm ws}$
  of diameter $O(\log(n))$, in comparison to the others.
\item Even on its intended graphs with heavy-tail degree
  distributions, \slashburn{} is consistently outperformed by \amd{}
  and \ned{}.
\item On networks with subcommunity structures, $G_{\rm PoK}$,
  $G_{\rm LFR}$, $G_{\rm football}$, $G_{\rm polblogs}$,
  $G_{\rm polbooks}$, $G_{\rm aps2020}$ and $G_{\rm Flickr}$, \amd{}
  is behind \rcm{}.
\end{inparaenum}
In summary, each scheme is limited to certain type(s) of
graphs.  By our analysis, each scheme is also short of reaching the
theoretically expected in its favorite type(s) of graphs.

\section{The new method: \vifps{} }
\label{sec:viFPS-description}

\begin{figure} 
  \centering
  \hspace{1em}
  \includegraphics[height = 0.45\linewidth]
  {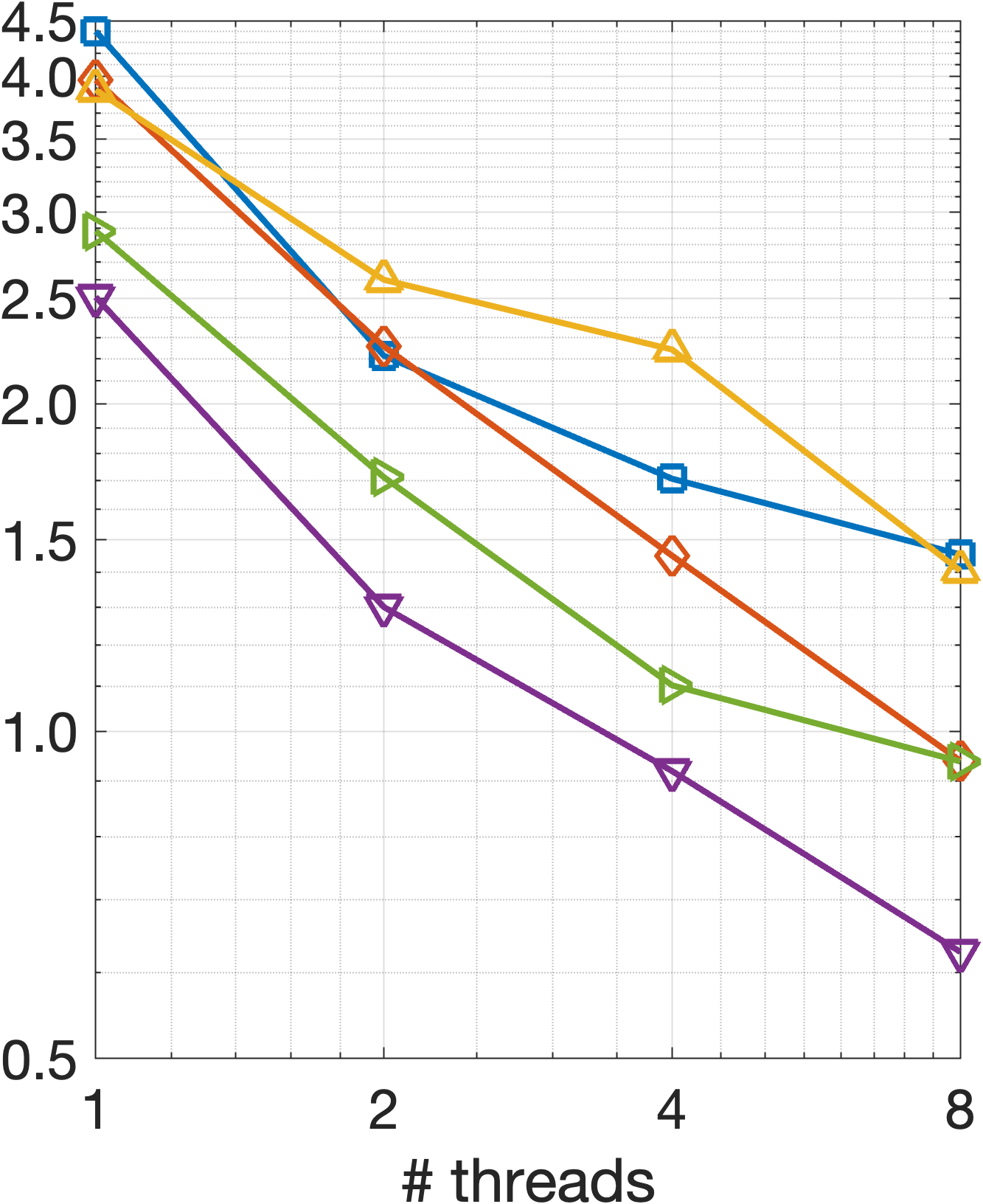}
  \hspace*{\fill}
  \includegraphics[height = 0.45\linewidth]
  {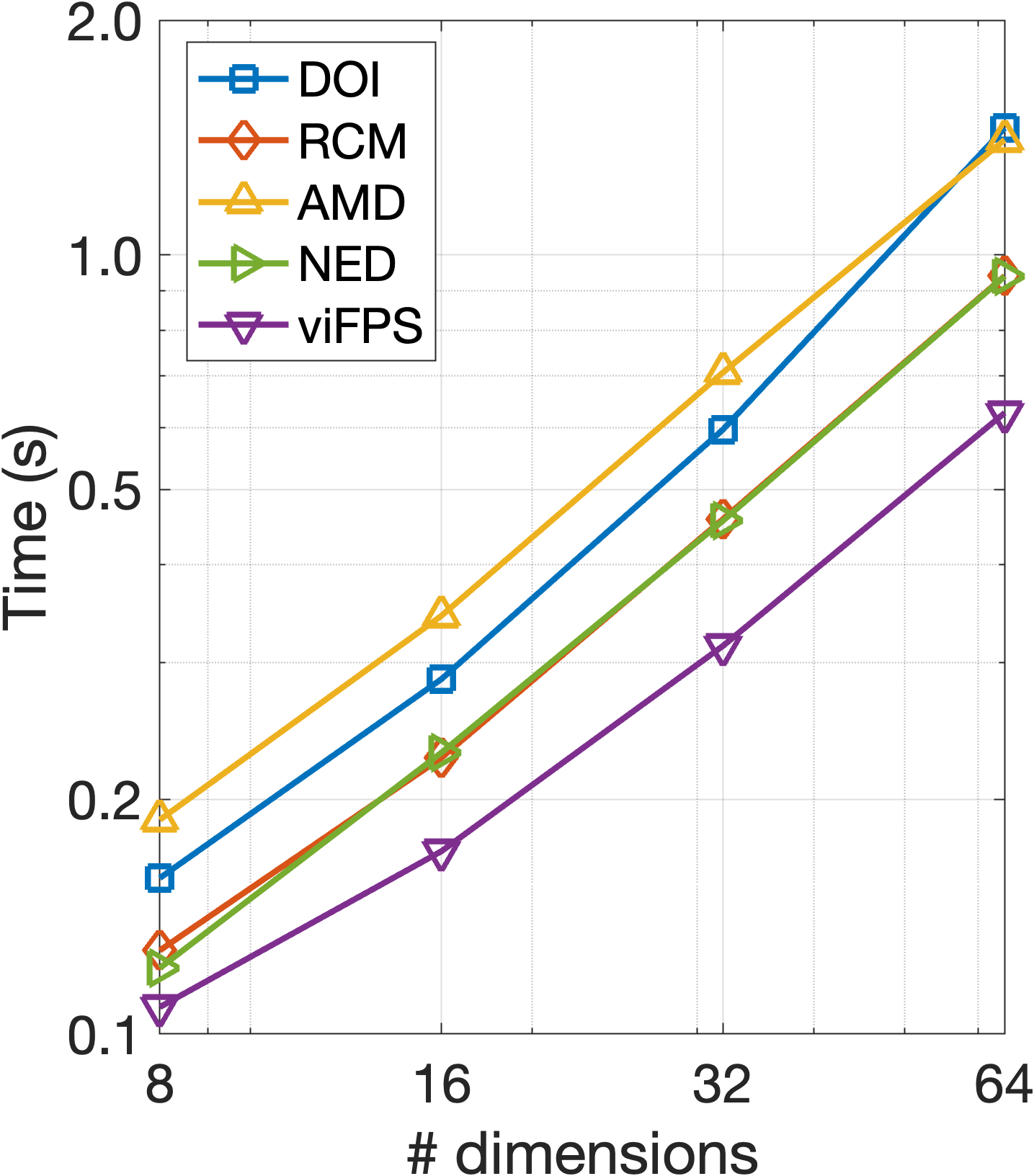}
  \hspace{2em}
\caption{The execution time in subspace iteration in responding to
    random-walk queries on the citation graph \aps{} with
    \num{5} different vertex orderings: \text{\sf DOI}, \rcm{},
    \amd{}, \ned{} and \vifps{}.  The measured time (sec) is for $10$
    iterations $x_{k+1}\! =\! A x_{k}$ with the adjacency matrix $A$
    and $d$ iterate vectors $x_{k}$, $d$ is the subspace dimension, on
    the {\sf Apple M2 Max} processor, using the package \textsf{SparseArrays} in
    \textsf{Julia}.  \textbf{Left:} The time variation with the number
    of threads $p\!\in\! \{1, 2, 4, 8\}$, at $d=64$.  \textbf{Right:}
    The time variation with the subspace dimension
    $d\!\in\! \{8,16, 32, 64\}$, at $p=8$.  \textbf{Observation:}
    \vifps{} shows better parallel scalability and cache utilization.
    \label{fig:aps-2020-subspace-iteration}
  }
  \vspace{-1.5em}
\end{figure}

We introduce a new method, \vifps{}.
In Table~\ref{tab:schemes-orders}, it demonstrates superior and
versatile performance in graph compression across diverse types of
networks and graphs. Figure~\ref{fig:aps-2020-subspace-iteration}
shows its additional benefits in improving the efficiency of subspace
iterations with a sparse matrix. We delineate its principled
properties and describe its simple procedure.

The development of \vifps{} originates from our better understanding
of the advantageous features and shortcomings of the existing methods.
\rcm{} keeps locally connected neighbors as close as possible.  \amd{}
adaptively, although implicitly, separates the globally shared
neighbors from the locally shared ones.  \slashburn{} is recursive but
inflexible in adjusting its removal of high-degree nodes through its
recursive division process, and it is unable to decouple the graph at
the weakest links as \fcutone{} does. Absent multi-resolution graph
partitions, \rcm{}, \fcutone{} and \amd{} are limited in extracting
substructures as \ned{} and \slashburn{} do. All but \fcutone{} are
combinatorial algorithms and sensitive to structural or random
variation.

We educe three principled properties for a versatile vertex ordering
method to acquire.
\begin{inparaenum}[(a)]
\item The persistent effectiveness in the presence of inevitable
  variations, structural or random, within an intended class or type
  of networks/graphs.
\item The generalizability to diverse types of networks/graphs.  This
  property warrants adaptability to the degree distribution. According
  to the \aal{} analysis in Section~\ref{sec:aal-analysis}, this also
  entails the algorithmic capability to decompose a graph into locally
  closely connected sub-communities detached from their commonly
  affiliated nodes, at multiple resolution levels.
  \item There is an explicit, easily interpretable, and computationally
  efficient approximation path, as with \fcutone{}, to the
  minimization of \mloggapa{} of (\ref{eq:mLogGapA}) over
  $\Pi(n)$.
\end{inparaenum}
Almost counterintuitively, making choices and modifications by these
properties, we arrive at a remarkably simple, recursive procedure.

Provided at input the adjacency matrix $A$ for a graph $G$,
a pair of Pareto ratios $(\text{rvol},\text{rminor})$, and a basecase
vertex set size $n_{\min}$, \vifps{} returns a vertex
permutation. The procedure is recursive, it takes the following steps
on the current graph $G$.
\begin{inparaenum}[(1)]
\item If in the basecase $|V(G)| \leq n_{\min}$, apply \amd{} to $G$
  and return the permutation.
\item \underline{The Pareto Split.}  If $\text{rvol}\%$ of the total
  volume $\sum_{i\in V}d(i)$ is held by $\text{rminor}\%$ or less of
  the nodes, split the minority nodes from the majority. Denote by
  $G_{\rm major}$ the graph induced by the majority nodes.
\item \underline{The Fiedler cuts.}  Apply \fcutone{} to every
  connected component of $G_{\rm major}$, followed by a recursive call
  of \vifps{} to every divided subgraph.
\item \underline{Aggregation}.  Return the permutation aggregated from
  the split and cuts.
\end{inparaenum}

The procedure extends to any digraph $G$, such as the
citation graph \aps{}, or a bipartite, via the embedding
matrix $[0,A; A^{\rm T},0]$. It returns both a row permutation and a
column perturbation.  We omit the rationale and nuance details due to
the document length limit.

The time complexity of \vifps{} scales with $c\, m\, \log_{2}(n)$,
where $m=|E|$, $n=|V|$, and $c$ is a modest constant, proportional to
the low dimension of a subspace iteration for obtaining the single
Fiedler vector of a sparse subgraph.

In the \vifps{} approach, the algebraic Fiedler cuts extract the
substructures, the statistical Pareto splits adapt to the degree
distributions of the divided subgraphs. If the split condition by the
ratio pair is not met on a subgraph, no split takes place. All splits
can be deactivated by setting the ratio pair as
$(\text{rvol},\text{rminor})=(100,1)$. Only in the ideal case of a
Pareto distribution, the split condition can be set by the single
Pareto scale parameter. We find that an estimate of the scale
parameter for a real-world network can be unreliable. For every
graph/network in Table~\ref{tab:schemes-orders}, we use a global split
condition $(\text{rvol},\text{rminor})\! =\! (20,4)$, which guides the
adaptive split in each divided subgraph. One can adjust or tune the
split control, globally or recursively, in an attempt to further
improve the compression, via parallel search or coordinated
search. The granularity for such parameter tuning is to be further
investigated in the cost-effectiveness aspect.

\section*{Acknowledgements}

We thank Mark Doyle and the American Physical Society (APS) for
providing the citation graph of APS publications up to the year 2020.
We thank an anonymous reviewer for helpful suggestions on further
clarification of the time-complexity scaling and the Pareto-split
condition setting.

\printbibliography 

\end{document}